\documentclass[iop]{emulateapj_mod}

\usepackage{epstopdf}

\usepackage{natbib}
\usepackage{hyperref}

\newcommand{\sj}{J1701}
\newcommand{\xte}{{\it RXTE}}
\newcommand{\sw}{{\it Swift}}
\newcommand{\cxo}{{\it Chandra}}
\newcommand{\xmm}{{\it XMM-Newton}}

\shorttitle{Rapid Cooling of the Neutron Star in XTE J1701--462}
\shortauthors{Fridriksson et al.}

\journalinfo{The Astrophysical Journal, 714:270--286, 2010 May 1}

\slugcomment{}

\begin{document}

\title{Rapid Cooling of the Neutron Star in the Quiescent Super-Eddington Transient~XTE~J1701--462}

\author{Joel K.\ Fridriksson\altaffilmark{1,2}, Jeroen Homan\altaffilmark{2}, Rudy Wijnands\altaffilmark{3}, Mariano M{\'e}ndez\altaffilmark{4}, Diego Altamirano\altaffilmark{3}, Edward M. Cackett\altaffilmark{5}, Edward F.\ Brown\altaffilmark{6}, Tomaso M.\ Belloni\altaffilmark{7}, Nathalie Degenaar\altaffilmark{3}, and Walter H.\ G.\ Lewin\altaffilmark{1,2}}

\altaffiltext{1}{Department of Physics, Massachusetts Institute of Technology, 77 Massachusetts Avenue, Cambridge, MA 02139; joelkf@mit.edu.}
\altaffiltext{2}{MIT Kavli Institute for Astrophysics and Space Research, 77 Massachusetts Avenue, Cambridge, MA 02139.}
\altaffiltext{3}{Astronomical Institute ``Anton Pannekoek,'' University of Amsterdam, Science Park 904, 1098 XH, Amsterdam, The Netherlands.}
\altaffiltext{4}{Kapteyn Astronomical Institute, University of Groningen, P.O. Box 800, 9700 AV, Groningen, The Netherlands.}
\altaffiltext{5}{Department of Astronomy, University of Michigan, 500 Church Street, Ann Arbor, MI 48109.}
\altaffiltext{6}{Department of Physics and Astronomy, National Superconducting Cyclotron Laboratory, and the Joint Institute for Nuclear Astrophysics, Michigan State University, East Lansing, MI 48824.}
\altaffiltext{7}{INAF -- Osservatorio Astronomico di Brera, Via E. Bianchi 46, I-23807 Merate (LC), Italy.}

\begin{abstract}

We present {\it Rossi X-Ray Timing Explorer} and \sw\ observations made during the final three weeks of the 2006--2007 outburst of the super-Eddington neutron star (NS) transient XTE~J1701--462, as well as \cxo\ and \xmm\ observations covering the first $\simeq$800 days of the subsequent quiescent phase. The source transitioned quickly from active accretion to quiescence, with the luminosity dropping by over 3 orders of magnitude in $\simeq$13 days. The spectra obtained during quiescence exhibit both a thermal component, presumed to originate in emission from the NS surface, and a non-thermal component of uncertain origin, which has shown large and irregular variability. We interpret the observed decay of the inferred effective surface temperature of the NS in quiescence as the cooling of the NS crust after having been heated and brought out of thermal equilibrium with the core during the outburst. The interpretation of the data is complicated by an apparent temporary increase in temperature $\simeq$220 days into quiescence, possibly due to an additional spurt of accretion. We derive an exponential decay timescale of $\simeq$$120^{+30}_{-20}$ days for the inferred temperature (excluding observations affected by the temporary increase). This short timescale indicates a highly conductive NS crust. Further observations are needed to confirm whether the crust is still slowly cooling or has already reached thermal equilibrium with the core at a surface temperature of $\simeq$125 eV. The latter would imply a high equilibrium bolometric thermal luminosity of $\simeq$$5\times10^{33}\textrm{ erg s}^{-1}$ for an assumed distance of 8.8 kpc.

\end{abstract}


\section{Introduction}\label{sec:intro}

Understanding the internal properties of neutron stars (NSs) remains one of the major unresolved problems in astrophysics. Efforts to constrain these properties most commonly focus on narrowing down the allowed regions in NS mass--radius diagrams, to thereby rule out some of the proposed equations of state for the matter inside the stars. An alternative approach is to observe the cooling of NSs \citep[for a review see][]{yakovlev2004}. Initially, this approach focused mainly on the long-term cooling of isolated NSs over the first $\sim$$10^6$ years after their birth, but it has in the past decade been extended to NSs reheated by transient accretion.

As matter from a binary companion is accreted onto the surface of a NS, matter already present is compressed further down into the crust to higher densities. This leads to heating due to nuclear reactions, so-called deep crustal heating \citep{brown1998,rutledge2002,haensel2008}. Most of the heat is produced by pycnonuclear reactions hundreds of meters below the surface and is spread throughout the star by heat conduction. Cooling takes place via neutrino emission from the interior and photon emission from the surface. In $\sim$$10^4$ years the NS enters a limit cycle in which heating during accretion episodes is on average balanced by cooling during and between outbursts \citep{colpi2001}. The temperature of the NS core is not expected to change appreciably after this, and its value depends on the long-term time-averaged mass accretion rate as well as on the efficiency of the cooling mechanisms at work, which in turn depend sensitively on the properties of the material inside the star. High-mass NSs are thought to potentially have much more powerful neutrino emission mechanisms active in their cores, compared to their low-mass counterparts \citep{yakovlev2003,yakovlev2004}; this is referred to as enhanced cooling, in contrast to the so-called standard cooling of the low-mass NSs.

The blackbody-like component often seen in spectra from NS low-mass X-ray binaries (NS-LMXBs) in quiescence is usually interpreted as thermal radiation from the surface of the NS. For a NS in thermal equilibrium, the temperature of the core can be inferred (via a model of the crust temperature profile) from the effective surface temperature \citep{yakovlev2004b}. A measurement of the thermal surface emission, in conjunction with information on the distance to the source and the average length, intensity, and recurrence time of accretion episodes, can thus possibly be used to constrain the properties of the material in the core of the NS.

In contrast to the core, the NS crust can have its temperature significantly altered, and be brought out of thermal equilibrium with the core, during single outbursts. In most NS X-ray transients, outbursts last from weeks to months, with periods of quiescence in between lasting from months to decades. The temperature of the crust is then expected to be raised only slightly during outbursts, and thermal equilibrium with the core will be quickly re-established (within days to weeks). However, in the so-called quasi-persistent transients, outbursts can last for several years or decades, in which case the crust is expected to cool down to equilibrium on a much longer timescale of several years \citep{rutledge2002}. It is therefore possible to monitor the cooling of such quasi-persistent transients with satellites such as \cxo\ or \xmm. The timescale of the cooling is dependent on the properties of the material in the crust, such as its thermal conductivity, and structures in the cooling curve can give information about the nature and location of heating sources in the crust \citep{brown2009}.

Since the advent of \cxo\ and \xmm, only a handful of NS transients have entered quiescence after long-duration (year or longer) outbursts. KS 1731--260 and MXB 1659--29 entered quiescence in 2001 after outbursts lasting around 12.5 and 2.5 years, respectively. Both sources were observed to cool down to a constant level over a period of a few years \citep{wijnands2001,wijnands2002a,wijnands2002b,rutledge2002,wijnands2003,wijnands2004a,wijnands2004b,cackett2006,cackett2008}, though a recent observation of KS 1731--260 at more than 3000 days post-outburst suggests it may still be cooling slowly (E.\ M.\ Cackett et al. 2010, in preparation). The observed cooling timescales were interpreted to imply a high thermal conductivity for the crust, in agreement with more recent findings from the fitting of theoretical models to the cooling curves \citep{shternin2007,brown2009}. In 2008, EXO 0748--676 entered quiescence after active accretion for over 24 years. \sw\ and \cxo\ observations of the source in the first half a year since the end of the outburst indicate very slow initial cooling \citep{degenaar2009}. In contrast to KS 1731--260 and MXB 1659--29, EXO 0748--676 has shown a significant non-thermal component in its spectra in addition to the thermal component. Such a non-thermal component has been seen for many quiescent NS-LMXBs. It is usually well fitted with a simple power law of photon index 1--2 and typically dominates the spectrum above a few keV \citep{campana1998}. A number of quiescent NS sources have spectra which are completely dominated by the power-law component and do not require a thermal component, e.g., the millisecond X-ray pulsar SAX J1808.4--3658 \citep{heinke2007} and the globular cluster source EXO 1745--248 \citep{wijnands2005}. The power-law component is common among millisecond X-ray pulsars \citep[see, e.g.,][]{campana2005}, but its origin is poorly understood. Suggested explanations include residual accretion, either onto the NS surface or onto the magnetosphere, and a shock from a pulsar wind \citep[see, e.g.,][]{campana1998}. We note that it has also been argued that low-level spherical accretion onto a NS surface can produce a spectrum with a thermal shape \citep{zampieri1995}.

\begin{figure}
\centerline{\includegraphics[width=8.5cm,trim=35 7 65 25,clip=true]{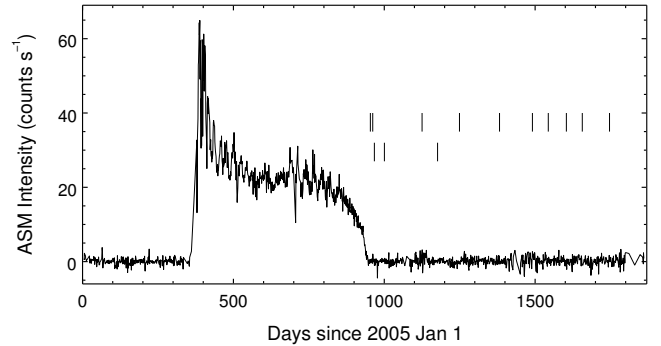}}
\caption{\xte\ ASM light curve of XTE J1701--462 showing the 2006--2007 outburst and the subsequent quiescent period. Data points represent 1 day averages. The upper row of vertical bars indicates the times of the ten \cxo\ observations made after the end of the outburst; the lower row indicates the times of the three \xmm\ observations. No other observations of XTE J1701--462 sensitive enough to detect the source have been made since the outburst ended.}\label{fig:outburst}
\end{figure}

\subsection{XTE J1701--462}\label{sec:1701}

XTE J1701--462 (hereafter \sj) was discovered with the All-Sky Monitor (ASM; \citealt{levine1996}) on board the {\it Rossi X-Ray Timing Explorer} (\xte) on 2006 January 18 \citep{remillard2006}, shortly after entering an outburst (see Figure~\ref{fig:outburst}). Re-analysis of earlier ASM data further constrained the start of the outburst to a date between 2005 December 27 and 2006 January 4 \citep{homan2007}. During the $\simeq$1.6-year-long outburst the source became one of the most luminous NS-LMXBs ever seen in the Galaxy, reaching a peak luminosity of $\simeq$$1.5$ $L_\mathrm{Edd}$, and it accreted at near-Eddington luminosities throughout most of the outburst \citep{lin2009a}. The source entered quiescence in early 2007 August (see Section~\ref{sec:cooling_curves} for a discussion of our definition of quiescence for this source). During the outburst the source was monitored on an almost daily basis with \xte. Spectral and timing analysis of the early phase of the outburst is presented in \citet{homan2007}, and \citet{lin2009a} give a detailed spectral analysis of the entire period of active accretion. In the early and most luminous phase of its outburst, \sj\ exhibited all spectral and timing characteristics typical of a Z source, and is the only transient NS-LMXB ever observed to do so. During the outburst the behavior of the source evolved through all spectral subclasses of low-magnetic-field NS-LMXBs \citep{hasinger1989}, starting as a Cyg-like Z source, then smoothly evolving into a Sco-like Z source \citep{kuulkers1997}, and finally into an atoll source (first a bright GX-like one and subsequently a weaker bursting one). This evolution will be discussed in detail in an upcoming paper (J.\ Homan et al.\ 2010, in preparation). The unique behavior of the source in conjunction with the dense coverage by \xte\ has made it possible to address long-standing questions regarding the role of mass accretion rate in causing these subclasses and the spectral states within each subclass \citep{lin2009a}. Toward the end of the outburst \sj\ exhibited three type I X-ray bursts, the latter two of which showed clear photospheric radius expansion. From these \citet{lin2009b} derive a best-estimate distance to the source of $8.8\pm1.3$ kpc, using an empirically determined Eddington luminosity for radius expansion bursts \citep{kuulkers2003}.

\sj\ provides a special test case for NS cooling. It accreted for a shorter time than the three cooling transients with long-duration outbursts mentioned above, but for a longer time than regular transients. Moreover, the level at which it accreted is higher than for any other NS transient observed. This source therefore allows new parameter space in NS cooling to be probed. The close monitoring of the source with \xte\ also makes it possible to get a good estimate for the total fluence of the outburst. This gives information about the total mass accreted and hence about the heat generated from crustal heating, a crucial input parameter for theoretical models of the cooling. Flux values derived from spectral fits to \xte\ data (spectra from 32 s time bins, with linear interpolation between data points; see Fig.\ 3 in \citealt{lin2009a}) imply a total bolometric energy output (corrected for absorption) during the outburst of $\simeq$$1.0\times10^{46}$ erg for an assumed distance of 8.8 kpc and system inclination of $70\degr$ (D. Lin 2009, private communication; see \citealt{lin2009a} for details on the spectral fitting). This value is likely to be uncertain by a factor of $\simeq$2--4 due to uncertainties in the distance and inclination of the system, as well as in the choice of a correct spectral model and the extrapolation of the model outside the \xte\ energy bandpass.

In this paper, we describe the end of \sj's outburst as the source transitioned from active accretion to quiescence and report on our monitoring of the subsequent cooling of the NS in quiescence.

\section{Data Analysis and Results}\label{sec:analysis}

\subsection{\textit{RXTE} Analysis}\label{sec:rxte_analysis}

A complete spectral analysis of \xte\ data from the outburst of \sj\ can be found in
\citet{lin2009a}. Here we present results from the analysis of \xte\ Proportional Counter Array (PCA; \citealt{jahoda2006}) data from 61 observations made during the last phase of the outburst and early quiescence (2007 July 17--August 29), with the main goal of obtaining flux values in the 0.5--10 keV band. For our analysis we only used data from Proportional Counter Unit 2 (PCU2). For each of the observations, a single PCU2 spectrum was extracted from {\tt Standard-2} mode data using HEASOFT, version 6.8. The spectra were corrected for dead time, background was subtracted, and a systematic error of 0.6\% was added to account for uncertainties in the PCA response. The three type I X-ray bursts discussed in Section~\ref{sec:1701} were removed from the data. A preliminary inspection of the light curves shows that the PCU2 count rate reached a nearly constant value of $\simeq$2 counts s$^{-1}$ after August 7. This flux is probably due to Galactic background emission, since {\it Swift} observations made in the last few days of the outburst (see Section~\ref{sec:swift_analysis}) indicate much lower fluxes than contemporaneous \xte\ observations. A spectrum of this residual emission was created by combining the spectra taken between August 8 and August 29. This spectrum was then subtracted from the spectra of observations taken before August 8; later observations were not considered further. We note that for the August 7 observation the residual emission represented $\simeq$2/3 of the 3.2--25 keV flux before subtraction.

\begin{figure}
\centerline{\includegraphics[width=8.5cm,trim=15 15 45 35,clip=true]{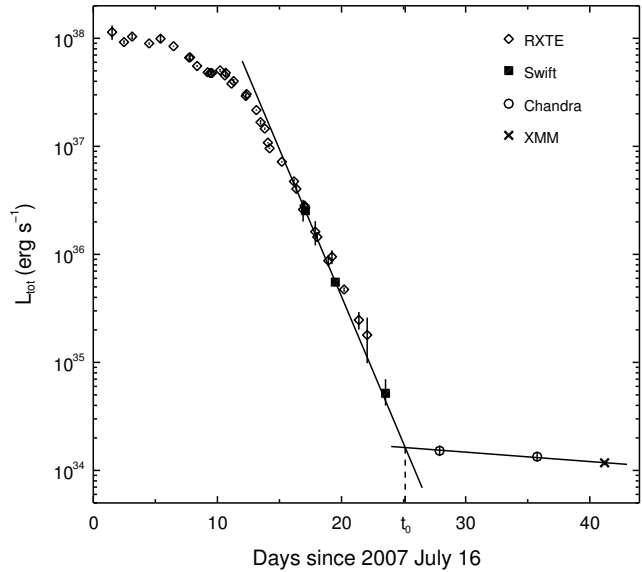}}
\caption{Total unabsorbed luminosity in the 0.5--10 keV band around the end of the outburst. The two lines are best-fit exponential decay curves for the three \sw\ observations, and the first three \cxo\ and \xmm\ observations. The intersection of these curves defines the end time of the outburst, $t_0$.}\label{fig:rxte}
\end{figure}

\begin{deluxetable*}{llccc}
\tablecaption{\sw, \cxo, and \xmm\ Observations of XTE J1701--462\label{obs_info}}
\tablehead{\colhead{Designation} & \colhead{Start Date} & \colhead{Instrument} & \colhead{ObsID}  & \colhead{Good Exp.\ Time (ks)}}
\startdata
Sw-1 & 2007 Aug 2 & {\it Swift} XRT & 00030383023 & 1.82\\
Sw-2 & 2007 Aug 4 & {\it Swift} XRT & 00030383024 & 1.93\\
Sw-3 & 2007 Aug 8 & {\it Swift} XRT & 00030383026 & 2.29\\
CXO-1 & 2007 Aug 12 & {\it Chandra} ACIS-S & 7513 & 4.69\\
CXO-2 & 2007 Aug 20 & {\it Chandra} ACIS-S & 7514 & 8.78\\
XMM-1 & 2007 Aug 26 & {\it XMM-Newton} EPIC & 0413390101 & 20.15 (MOS1)\\
 & & & & 19.27 (MOS2)\\
 & & & & 9.51(pn)\\
XMM-2 & 2007 Sep 28 & {\it XMM-Newton} EPIC & 0413390201 & 22.02 (MOS1)\\
 & & & & 23.12 (MOS2)\\
 & & & & 12.36 (pn)\\
CXO-3 & 2008 Jan 31 & {\it Chandra} ACIS-S & 7515 & 19.91\\
XMM-3 & 2008 Mar 22 & {\it XMM-Newton} EPIC & 0510990101 & 30.36 (MOS1)\\
 & & & & 31.05 (MOS2)\\
 & & & & 21.65 (pn)\\
CXO-4 & 2008 Jun 3 & {\it Chandra} ACIS-S & 7516 & 27.37\\
CXO-5 & 2008 Oct 13 & {\it Chandra} ACIS-S & 7517 & 39.92\\
CXO-6 & 2009 Jan 30 & {\it Chandra} ACIS-S & 10063 & 50.06\\
CXO-7a & 2009 Mar 23 & {\it Chandra} ACIS-S & 10064 & 13.26\\
CXO-7b & 2009 Mar 23 & {\it Chandra} ACIS-S & 10891 & 9.02\\
CXO-7c & 2009 Mar 24 & {\it Chandra} ACIS-S & 10889 & 11.88\\
CXO-7d & 2009 Mar 25 & {\it Chandra} ACIS-S & 10890 & 19.57\\
CXO-8 & 2009 May 23 & {\it Chandra} ACIS-S & 10065 & 62.31\\
CXO-9 & 2009 July 16 & {\it Chandra} ACIS-S & 10066 & 65.36\\
CXO-10a & 2009 Oct 13 & {\it Chandra} ACIS-S & 10067 & 43.56\\
CXO-10b & 2009 Oct 13 & {\it Chandra} ACIS-S & 12006 & 25.80
\enddata
\end{deluxetable*}

The spectra were fitted in the 3.2--25 keV range with XSPEC \citep{arnaud1996}, version 12.5.1. For the first observations we used a model consisting of a multicolor disk blackbody ({\tt diskbb} in XSPEC), a blackbody ({\tt bbody}), a power law ({\tt powerlaw}), and a Gaussian emission line ({\tt gauss}). These components were modified by a photoelectric absorption component ({\tt phabs}) with the column density fixed at $N_H=1.93\times10^{22}\textrm{ cm}^{-2}$ (as determined from our fits to the \cxo\ and \xmm\ spectra; see Section~\ref{sec:fitting_results}). Following \citet{lin2009a} we fixed the energy and width of the Gaussian at 6.5 keV and 0.3 keV, respectively, and constrained the power-law index to be lower than or equal to 2.5. In addition to these constraints, we also fixed the normalization of the {\tt diskbb} component at a value of 17.0 in XSPEC; this component otherwise becomes poorly constrained during the decay, leading to large uncertainties when extrapolating the spectral model below the \xte\ energy bandpass. The value of 17.0 is an average of values found for the disk normalization in the atoll phase of the outburst (when the radius of the disk was approximately constant). The {\tt diskbb} and {\tt bbody} components were statistically no longer needed in spectra taken after approximately July 29 08:00 and August 4 12:00 UT, respectively, and were dropped from the spectral model after those times. The best-fit model was extrapolated down to 0.5 keV (well below the lower-energy boundary of the PCA) to obtain 0.5--10 keV unabsorbed fluxes. Such an extrapolation can give rise to significant systematic errors in the flux values, due to uncertainty in the choice of a proper spectral model. This is on top of possible error associated with the subtraction of the background Galactic emission. However, the good agreement between the \xte\ and contemporaneous \sw\ observations (see Figure~\ref{fig:rxte}) indicates that the \xte\ flux values do not suffer from large systematic errors. The derived fluxes for the first \sw\ observation and a simultaneous \xte\ observation differ by only $\simeq$6\% and agree within the uncertainties due to counting statistics. We note that since this paper is primarily concerned with the quiescent phase of the source, none of our results are affected by the values of the \xte\ fluxes. When calculating the luminosities for those \xte\ observations that included a disk blackbody component in their spectra (i.e., observations in the first $\simeq$13 days in Figure~\ref{fig:rxte}), a disk inclination of $70\degr$ was assumed \citep{lin2009a}. Possible error in these luminosities due to uncertainty in the assumed inclination is probably at most a factor of 2 for the earliest observations; less for the later ones, since the contribution of the disk blackbody component to the total flux gradually decreases. The \xte\ errors plotted in Figure~\ref{fig:rxte} are the same fractional errors as those derived for the absorbed 3.5--10 keV flux in XSPEC, and only take into account uncertainty due to counting statistics (in addition to the assumed 0.6\% systematic error arising from uncertainty in the PCA response). We note here that all errors quoted in this paper correspond to $1\sigma$ Gaussian (68.3\%) confidence.

\subsection{\textit{Swift} Analysis}\label{sec:swift_analysis}

\sj\ was observed 3 times with the \sw\ X-Ray Telescope (XRT; \citealt{burrows2005}) in the last few days of the outburst (see Table~\ref{obs_info}). All three observations were made with the detector in the Photon Counting mode. We analyzed the data using HEASOFT (ver. 6.8), and the latest calibration files available at the time of the analysis. Starting with the Level 1 raw event files, the data were processed and screened using the {\tt xrtpipeline} task with the default parameter settings, as described in The \sw\ XRT Data Reduction Guide.\footnote{Available at \url{http://swift.gsfc.nasa.gov/docs/swift/analysis/xrt\_swguide\_v1\_2.pdf}.} Additionally, a short period with an increased background count rate was filtered out in the third observation. We extracted source and background spectra from the Level 2 screened event files using Xselect. The first observation suffered from moderate pile-up and therefore a core of radius $12\arcsec$ was excised from an extraction circle of radius $70\arcsec$. The appropriate excision radius was determined by comparing the observed point-spread function (PSF) to the empirically determined (pile-up free) one \citep{moretti2005}. For the second and third observations, source spectra were extracted from circular regions of radii $47\arcsec$ and $35\arcsec$, respectively. Background spectra were in all three cases extracted from a source-free circular region of radius $200\arcsec$. We used a standard redistribution matrix file (RMF) from the calibration database and created observation-specific ``empirical'' (i.e., adjusted to fit the Crab spectrum; see the Data Reduction Guide) ancillary response files (ARFs) using the {\tt xrtmkarf} task. The ARFs were corrected for counts missed by the finite extraction regions (as were our \cxo\ and \xmm\ ARFs).

The spectra were fitted in the 0.2--10 keV band with XSPEC (ver. 12.5.1). We binned the spectra from the first two observations into groups with a minimum of 25 counts. Due to the low number of source counts in the third observation (52), that spectrum was left unbinned and fitted using the W statistic (a modification of the C statistic that allows for background subtraction; see \citealt{wachter1979} and the XSPEC User's Guide\footnote{Available at \url{http://heasarc.gsfc.nasa.gov/docs/xanadu/xspec/manual/manual.html}.}). We fixed the absorption column at the value of $1.93\times10^{22}\textrm{ cm}^{-2}$ derived from our fits to the \cxo\ and \xmm\ spectra. All three \sw\ spectra were adequately fitted with a simple absorbed power law. For none of the observations did the addition of a thermal component to the spectral model improve the fit. The best-fit values for the power-law index are $1.68\pm0.06$ for the first observation, $1.92\pm0.09$ for the second one, and $2.8\pm0.4$ for the third.

\subsection{\textit{Chandra} Observations}\label{sec:chandra_analysis}

All the {\it Chandra} observations made after \sj\ entered quiescence (see Table~\ref{obs_info}) used ACIS-S imaging \citep{garmire2003} in the Timed Exposure mode, with the source located at the nominal aimpoint on the S3 chip. The detector was operated either in full-frame (frame time of $\simeq$3.2 s) or 1/8-subarray mode (frame time of $\simeq$0.4 s), using the Faint or Very Faint telemetry format. We analyzed the observations using the CIAO software \citep{fruscione2006}, version 4.2 (CALDB ver.\ 4.2.0), and with ACIS Extract\footnote{The ACIS Extract software package and User's Guide are available at
  \url{http://www.astro.psu.edu/xray/acis/acis_analysis.html}.} \citep{broos2010}, version 2010-01-07. The data were processed following the standard ACIS data preparation procedure recommended by the \cxo\ X-Ray Center.\footnote{See \url{http://cxc.harvard.edu/ciao/guides/acis\_data.html}.} We checked the observations for possible periods of background flaring, using background light curves from the entire S1 chip (for the full-frame observations) or the entire active area of the S3 chip (for the subarray observations). No such periods were found, except for a short spike in Observation ID 10066 which was excluded from the data. We performed further analysis with the help of ACIS Extract. Source spectra were extracted from polygon-shaped regions modeled on the \cxo\ ACIS PSF using the MARX ray-trace simulator. The extraction regions had a PSF enclosed energy fraction of $\simeq$0.97 (for a photon energy of $\simeq$1.5 keV) and a radius of $\simeq$$1\farcs9$. Background spectra were extracted from source-free circular annuli centered on the source location, with an inner radius of 1.5 times the radius that encloses 99\% of the PSF. We chose the outer radius so that the background region had at least 4 times the exposure-corrected area of the source extraction region, and a minimum of 100 counts (the latter condition in all cases leading to a much higher ratio than 4 between the areas of the background and source regions). Response files were created using the {\tt mkacisrmf} and {\tt mkarf} tools in CIAO.

\subsection{\textit{XMM-Newton} Observations}\label{sec:xmm_analysis}

The three \xmm\ observations made of \sj\ in quiescence (see Table~\ref{obs_info}) used EPIC imaging \citep{turner2001,struder2001} with the MOS detectors operated in the small-window or full-frame mode, and the pn detector in the extended-full-frame mode. The thin filter was used in all cases. We analyzed the data using the SAS software, version 9.0.0, and the latest calibration files available at the time of the analysis. Starting with the original Observation Data Files, the data were reprocessed using the {\tt emproc} and {\tt epproc} tasks. Significant portions of the exposures had to be excluded due to periods of increased particle background. These periods were identified by constructing light curves from the entire area of the detectors, using only single events (PATTERN=0) and high energy photons (larger than 10 keV for MOS and between 10 and 12 keV for pn). Intervals with count rates above a certain level (0.18--0.24 counts s$^{-1}$ for MOS, 0.36--0.40 counts s$^{-1}$ for pn) were then excluded. An additional time cut was made to better filter out a large flare at the end of the third observation. We used the {\tt evselect} task to extract spectra, after utilizing the {\tt eregionanalyse} task to optimize the circular source extraction region in each case (i.e., center the region on the centroid of the counts distribution and find the radius that maximizes the ratio of source counts to background counts). Extraction radii in the range $25\arcsec$--$45\arcsec$ were used, giving PSF enclosed energy fractions in the range $\simeq$0.83--0.87. Background spectra were extracted from source-free circular regions with roughly 4 times the area of the source extraction regions. Following the recommendations in \citet{guainazzi2010}, we extracted background spectra for the MOS detectors from another region on the same CCD as the source was on, away from source counts; for the pn detector, the background region was on an adjacent CCD to the one the source was on, at a similar distance to the readout node as the source region. The event selection criteria used were PATTERN=0--12 and FLAG=\#XMMEA\_EM for the MOS detectors, and PATTERN=0--4 and FLAG=0 for the pn detector. We created both the MOS and pn spectra with a resolution of 30 eV; this ensured that the grouped spectra used for fitting (see Section~\ref{sec:fitting_results}) would not oversample the resolution of the detectors by more than a factor of 2--3. Response files were created using the {\tt rmfgen} and {\tt arfgen} tasks.

\subsection{Fitting of \textit{Chandra} and \textit{XMM-Newton} Spectra}\label{sec:spectral_fitting}

\subsubsection{Spectral Models}\label{sec:spectral_models}

The \cxo\ and \xmm\ spectra were modeled with two source components, a NS atmosphere model (thermal component) and a simple power-law model (non-thermal component), along with an overall photoelectric absorption component. For the latter, we used the {\tt phabs} model in XSPEC with the default cross sections and relative abundances.

\begin{deluxetable*}{lcccccc}
\tablecaption{Selected Spectral Fit Parameters\label{spectral_param}}
\tablehead{\colhead{Fit} & \colhead{$D$ (kpc)} & \colhead{$R_\mathrm{ns}$ (km)} & \colhead{$M_\mathrm{ns}$ ($M_\sun$)} & \colhead{$N_\mathrm{H}$ ($10^{22}\textrm{ cm}^{-2}$)} & \colhead{$\alpha$\tablenotemark{a}} & \colhead{$\alpha'$\tablenotemark{b}}}
\startdata
1 & (8.8) & (10) & (1.4) & $1.93\pm0.02$ & $1.93\pm0.20$ & $1.35\pm0.08$ \\
2 & (7.5) & (10) & (1.4) & $1.98\pm0.03$ & $2.04\pm0.19$ & $1.37\pm0.08$ \\
3 & (10.1) & (10) & (1.4) & $1.88\pm0.02$ & $1.83\pm0.20$ & $1.34\pm0.08$ \\
4 & $7.3\pm1.4$ & (10) & (1.4) & $1.99\pm0.07$ & $2.06\pm0.23$ & $1.37\pm0.08$ \\
5 & (8.8) & $11.7\pm2.0$ & (1.4) & $1.98\pm0.06$ & $2.02\pm0.22$ & $1.36\pm0.08$ \\
6 & (8.8) & (10) & $1.6\pm0.4$ & $1.95\pm0.06$ & $1.98\pm0.21$ & $1.36\pm0.08$ \\
7 & (8.8) & $11.6\pm2.6$ & $1.5\pm0.6$ & $1.98\pm0.07$ & $2.03\pm0.24$ & $1.36\pm0.08$ \\
8 & (8.8) & (10) & (1.4) & $1.95\pm0.02$ & $(1.3)$ & $1.34\pm0.08$ \\
9 & (8.8) & (10) & (1.4) & $1.92\pm0.02$ & $(2.5)$ & $1.35\pm0.08$
\enddata
\tablecomments{All \cxo\ and \xmm\ observations were fitted simultaneously. All the fits have a reduced $\chi^2$ of 1.11--1.13 for 515--518 dof. Numbers in parentheses indicate parameters fixed during fitting. Errors quoted are at the $1\sigma$ Gaussian (68.3\%) confidence level. --- $^\mathrm{a}\textrm{ }$Combined (tied) power-law index for all observations except the third \xmm\ observation (XMM-3). $^\mathrm{b}\textrm{ }$Power-law index for XMM-3.}
\end{deluxetable*}

NS atmosphere spectra have a blackbody-like shape, but with an important difference to classic blackbodies being a shift in the peak of the emitted radiation to higher photon energies and a harder high-energy tail. This is due to the strong energy dependence of free--free absorption in the NS atmosphere leading to higher energy photons coming from deeper and hotter layers in the atmosphere \citep[see, e.g.,][]{zavlin2002}. Fitting spectra from NS atmospheres with a classic blackbody model therefore leads to an overestimate of the effective temperature, and to an underestimate of the radius of the emitting region by as much as an order of magnitude \citep{rutledge1999}. We used the NS atmosphere model {\tt nsatmos}, available in XSPEC \citep[for detailed information on the model, see][]{heinke2006}. This model has five fitting parameters: the unredshifted effective surface temperature of the NS ($T_\mathrm{eff}$), the NS mass ($M_\mathrm{ns}$), the true NS radius ($R_\mathrm{ns}$), the distance to the NS ($D$), and an additional normalization parameter representing the fraction of the NS surface emitting radiation. For given values of $M_\mathrm{ns}$ and $R_\mathrm{ns}$ one can, from $T_\mathrm{eff}$, calculate the more often quoted redshifted effective temperature as measured by an observer at infinity as $T_\mathrm{eff}^\infty=T_\mathrm{eff}/(1+z)$, where $1+z=(1-R_\mathrm{S}/R_\mathrm{ns})^{-1/2}$ is the usual gravitational redshift factor, with $R_\mathrm{S}=2GM_\mathrm{ns}/c^2$ being the Schwarzschild radius. One can also calculate the observed NS radius at infinity as $R_\mathrm{ns}^\infty=R_\mathrm{ns}(1+z)$. The canonical values for the NS parameters of $M_\mathrm{ns}=1.4\textrm{ }M_\sun$ and $R_\mathrm{ns}=10\textrm{ km}$ give a redshift factor of $1+z\simeq1.306$.

The {\tt nsatmos} model makes several simplifying assumptions, including the following. (1) Magnetic fields are assumed to have negligible effects on the spectrum from the NS atmosphere (i.e., $B\lesssim10^8$--$10^9$~G). Given that J1701 has shown behavior typical of low-magnetic-field NS-LMXBs, with no X-ray pulsations having been detected, this should be a reasonably good assumption. (2) The NS atmosphere is assumed to be composed of pure hydrogen. Since heavier elements are expected to settle out of the atmosphere on a timescale of $\sim$10 s \citep{romani1987,bildsten1992}, this should be true if the accretion rate is below $\sim$$10^{-13}\textrm{ }M_\sun\textrm{ yr}^{-1}$ \citep{brown1998}. If the non-thermal component seen in the spectra from J1701 is due to a residual accretion flow which is radiating efficiently, then the accretion rate is likely $\sim$$10^{-14}$--$10^{-12}\textrm{ }M_\sun\textrm{ yr}^{-1}$ for our observations. \citet{heinke2006} note that atmospheres with iron or with solar abundances would be easily identifiable by their different spectral shapes, whereas small departures from a pure hydrogen atmosphere may go unnoticed and affect results. It is therefore conceivable that our results may be affected by some contamination of the hydrogen atmosphere, although the fact that our spectra are in general well fitted with {\tt nsatmos} (with an implied emission area consistent with what is expected from a NS) suggests that this is not a serious problem. (3) Finally, it is assumed that the hydrogen is completely ionized, and Comptonization is ignored. This limits the validity of the code to a temperature range of $3\times10^5\textrm{ K}\lesssim T_\mathrm{eff}\lesssim3\times10^6\textrm{ K}$ \citep{heinke2006}, corresponding to $20\textrm{ eV}\lesssim kT_\mathrm{eff}^\infty\lesssim200\textrm{ eV}$ for typical values of the NS parameters. The derived effective temperatures for J1701 fall well within this range.

In addition to {\tt nsatmos}, two other spectral models for NS atmospheres with negligible magnetic fields are available in XSPEC, {\tt nsa} and {\tt nsagrav}. The {\tt nsatmos} and {\tt nsagrav} models allow for variations in surface gravity corresponding to different values of the NS radius and mass. In contrast, {\tt nsa} uses a single fixed surface gravity value. Comparisons of different NS atmosphere models have shown that taking variations in surface gravity into account can be important when allowing the NS mass and/or radius to vary \citep{heinke2006,webb2007}; {\tt nsatmos} and {\tt nsagrav} are therefore to be preferred. We did a cursory comparison of our spectral fitting results when using {\tt nsatmos} and {\tt nsagrav}, and found them to be largely equivalent. In what follows we only report results obtained using {\tt nsatmos}.

We model the non-thermal component with a simple power law (using the {\tt pegpwrlw} model in XSPEC, whose normalization is pegged to equal the energy flux of the model for a chosen energy range). We note that this is not a physically motivated choice (the origin of this component is uncertain), but is simply based on the fact that non-thermal components in quiescent NS-LMXB spectra have often been successfully modeled by power laws in the past. It is, however, very possible that our results may be somewhat affected by the power-law model not properly representing the true shape of the non-thermal component. To get some indication of how our results might be affected by the choice of a model for the non-thermal component, we also performed a fit where the power-law model was replaced with the {\tt simpl} model in XSPEC (see discussion in Section~\ref{sec:tests}).

\begin{deluxetable*}{lrccccc}
\tablecaption{Derived Temperatures and Fluxes from {\it Chandra} and {\it XMM-Newton} Observations\label{results}}
\tablehead{\colhead{Observation} & \colhead{$t-t_0$\tablenotemark{a}} & \colhead{$kT^{\infty}_\mathrm{eff}$\tablenotemark{b}} &  \colhead{$F_\mathrm{bol}$\tablenotemark{c}}  & \colhead{$F_\mathrm{pl}$\tablenotemark{d}} & \colhead{$L_\mathrm{tot}$\tablenotemark{e}}\\
\colhead{} & \colhead{(days)} & \colhead{(eV)} & \colhead{($10^{-13}\textrm{ erg s}^{-1}\textrm{ cm}^{-2}$)} & \colhead{($10^{-13}\textrm{ erg s}^{-1}\textrm{ cm}^{-2}$)} & \colhead{($10^{33}\textrm{ erg s}^{-1}$)}
}
\startdata
CXO-1  & 2.77 & $164.2\pm3.6$         & $17.2\pm1.5$          & $\phn1.5\pm1.1$            & $15.2\pm1.7$\\
CXO-2  & 10.63 & $159.5\pm2.5$         & $15.4\pm1.0$          & $\phn1.3\pm0.7$            & $13.4\pm1.2$\\
XMM-1 & 16.06 & $156.8\pm1.3$         & $14.3\pm0.5$          & $\phn0.5\pm0.3$            & $11.8\pm0.5$\\
XMM-2 & 49.31 & $150.0\pm1.2$         & $12.0\pm0.4$          & $\phn0.8\pm0.3$            & $10.1\pm0.5$\\
CXO-3  & 174.15 & $129.1\pm4.7$ 	& $\phn6.6\pm1.0$ 		  & $\phn4.8\pm0.9$ 	     & $9.3^{+0.6}_{-0.8}$\\
XMM-3 & 225.54 & $159.3\pm2.0$         & $15.3\pm0.8$          & $15.1\pm0.6$          & $26.1\pm0.5$\\
CXO-4  & 298.12 & $136.0\pm2.0$         & $\phn8.1\pm0.5$            & $\phn1.4\pm0.3$            & $\phn7.3\pm0.5$\\
CXO-5  & 430.89 & $126.3\pm3.1$ 	& $\phn6.0\pm0.6$ 		   & $\phn3.1\pm0.5$ 	     & $7.2^{+0.4}_{-0.5}$\\
CXO-6  & 539.90 & $125.4\pm1.5$ 	& $\phn5.8\pm0.3$		   & $\phn0.5\pm0.2$ 	     & $\phn4.7\pm0.3$\\
CXO-7\tablenotemark{f}  & 592.50 & $129.6\pm2.2$ & $\phn6.7\pm0.5$& $\phn2.0\pm0.4$    & $6.8^{+0.3}_{-0.5}$\\
CXO-8  & 652.44 & $124.0\pm2.2$ 	& $\phn5.6\pm0.4$		   & $\phn1.8\pm0.3$ 	     & $5.7^{+0.3}_{-0.4}$\\
CXO-9  & 705.20 & $123.9\pm2.0$ 	& $\phn5.6\pm0.4$		   & $\phn1.5\pm0.3$ 	     & $5.4^{+0.3}_{-0.4}$\\
CXO-10\tablenotemark{g}  & 795.45 & $124.1\pm1.7$ & $\phn5.6\pm0.3$ & $\phn1.0\pm0.2$  & $\phn5.0\pm0.3$
\enddata
\tablecomments{Values were obtained from a simultaneous spectral fit to all observations, with the NS mass, radius, and distance fixed at 1.4 $M_\sun$, 10 km, and 8.8 kpc (fit 1 in Table~\ref{spectral_param}). Errors quoted are at the $1\sigma$ Gaussian (68.3\%) confidence level. --- $^\mathrm{a}\textrm{ }$Time of mid-observation; $t_0$ is MJD 54322.13. $^\mathrm{b}\textrm{ }$Effective surface temperature of the NS as seen by an observer at infinity. $^\mathrm{c}\textrm{ }$Unabsorbed bolometric flux of the thermal ({\tt nsatmos}) component. $^\mathrm{d}\textrm{ }$Unabsorbed 0.5--10 keV flux of the non-thermal (power-law) component. $^\mathrm{e}\textrm{ }$Unabsorbed total luminosity in the 0.5--10 keV band. $^\mathrm{f}\textrm{ }$Values derived with parameters for the CXO-7[a-d] spectra tied. $^\mathrm{g}\textrm{ }$Values derived with parameters for the CXO-10[a-b] spectra tied.}
\end{deluxetable*}

\subsubsection{Fitting Results}\label{sec:fitting_results}

We fitted the \cxo\ and \xmm\ spectra in the 0.5--10 keV band with XSPEC (ver. 12.5.1), after binning the spectra into groups with a minimum signal-to-noise ratio of 4 using the ACIS Extract tool {\tt ae\_group\_spectrum} (resulting in an average number of counts per group of $\simeq$25--30 for the \cxo\ spectra and $\simeq$35--40 for the \xmm\ spectra). The spectra from all 13 \cxo\ and \xmm\ observations were fitted simultaneously. The seventh \cxo\ observation (CXO-7) consisted of four separate exposures taken over a period of approximately 3 days (see Table~\ref{obs_info}). In the main spectral fit, used to derive temperatures and fluxes, we tied all the parameters for these four spectra, thereby effectively treating the exposures as a single observation. However, we also performed a fit without the parameters being tied (see Section~\ref{sec:tests}). The tenth and most recent \cxo\ observation consisted of two separate exposures taken over a $\simeq$27 hr period; this observation was treated in the same manner as CXO-7. For each \xmm\ observation, all fitting parameters were tied between the spectra from the three EPIC detectors (MOS1, MOS2, and pn). Several parameters were tied between all 13 observations: the absorption column, the NS mass, radius, and distance, and the fraction of the NS surface emitting. Additionally, due to the limited statistics of our spectra and the fact that we are mainly interested in constraining the values for the effective temperature (as well as the flux contribution from the non-thermal component), most of the {\tt nsatmos} parameters were usually fixed at a certain value and not allowed to vary during the fitting. (1) The value of the distance parameter was fixed at the best-estimate value of 8.8 kpc (see \S~\ref{sec:1701}) in the main fit, although we also performed fits where the distance was allowed to vary or was fixed at 7.5 kpc or 10.1 kpc. (2) Likewise, the NS mass and radius were fixed at the canonical values of 1.4 $M_\sun$ and 10 km in the main fit, although we also experimented with allowing them to vary (see below). (3) Finally, since we assume that the entire surface of the NS is emitting thermal radiation during quiescence, the fraction of the NS surface emitting was always fixed at 1.

\begin{figure}
\centerline{\includegraphics[width=6.3cm,trim=0 0 0 0,angle=-90,clip=true]{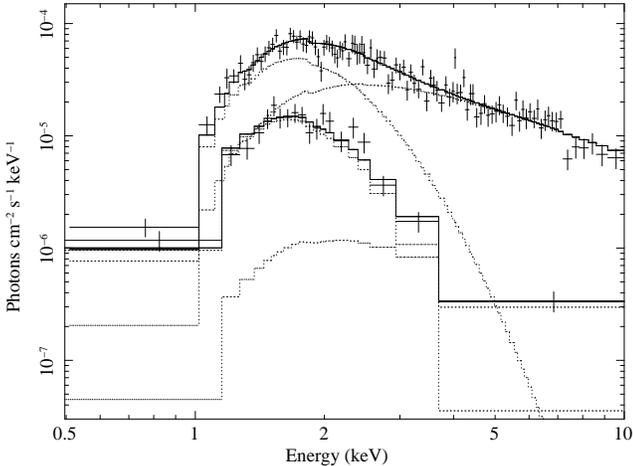}}
\caption{Fits to the pn spectrum for the third {\it XMM-Newton} observation (upper) and the spectrum for the sixth {\it Chandra} observation (lower). Also shown (dotted lines) are the contributions from the two spectral components: the thermal component, which dominates at lower energies, and the non-thermal component, which dominates at higher energies. The upper spectrum has the largest non-thermal component of all the observations, whereas the lower spectrum has a non-thermal component among the smallest seen.}\label{fig:spectra}
\end{figure}

Although we have no reason to believe that the shape of the non-thermal component should necessarily be the same in all the observations (as mentioned above, its origin is highly uncertain), we do not have enough counts to allow the power-law index to vary freely for each observation. Doing so leads to very poorly constrained values for the index in most cases and a very unstable fit. An exception is the third \xmm\ observation (XMM-3), which has a much larger non-thermal component than the other observations. In the main fit, we therefore tie the power-law index between all the observations except XMM-3. Tying the XMM-3 power-law index to the other observations would lead to it completely dominating the fit for that parameter. To gauge the effect of this tying on the derived temperatures and fluxes we also performed fits where the index was fixed at 1.3 and 2.5 for all observations except XMM-3 (see discussion in Section~\ref{sec:tests}). The only parameters allowed to vary independently for each observation in the main fit were the effective NS surface temperature and the power-law normalization. We note that the values obtained for the power-law indices when allowing them to vary freely for each observation are nearly all consistent (within their large error bars) with the single value obtained when tying the index as described above. We also note that the power-law component is clearly required for nearly all the individual observations (and not just XMM-3) to get an acceptable fit, and is included in all observations for consistency.

\begin{figure}
\centerline{\includegraphics[width=8.5cm,trim=25 52 45 65,clip=true]{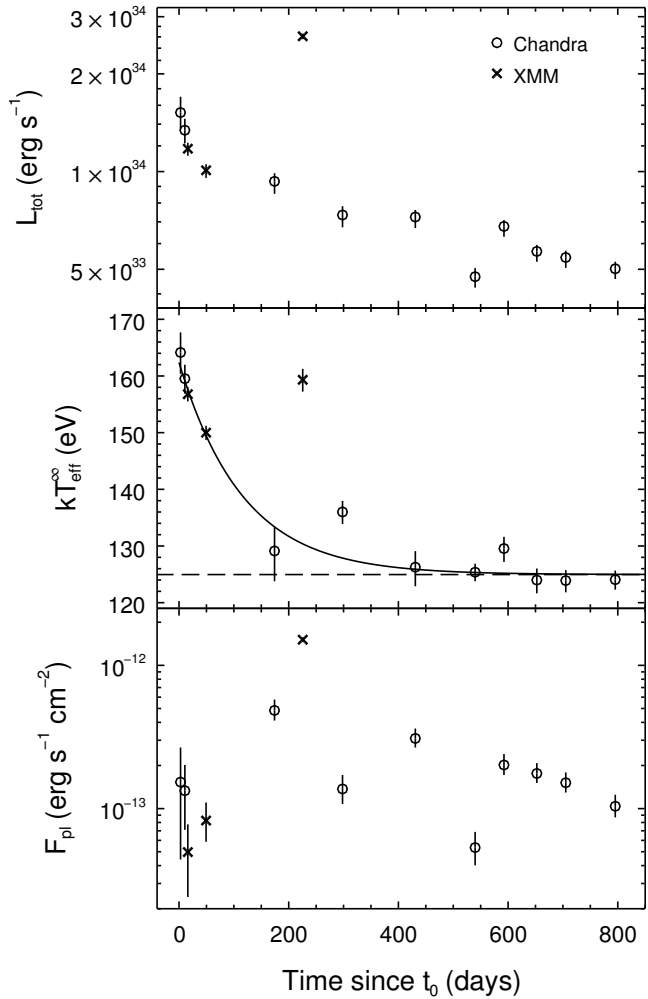}}
\caption{Total unabsorbed luminosity in the 0.5--10 keV band (top panel), redshifted effective NS surface temperature (middle panel), and unabsorbed power-law flux in the 0.5--10 keV band (bottom panel) during quiescence. The solid curve in the temperature panel is the best-fit exponential decay cooling curve (with the sixth and seventh data points excluded from the fit), and the dashed line represents the best-fit constant offset to the decay.}\label{fig:lum+cooling+pl}
\end{figure}

In Table~\ref{spectral_param}, we show the results of spectral fits with different combinations of fixed and free NS parameters. We do not show the parameters that were allowed to vary freely for each individual observation (i.e., the effective temperatures and power-law normalizations); those would add 26 additional numbers for each fit to the table. Fit 1 in Table~\ref{spectral_param} is the main fit that was used to derive temperatures and fluxes used in most of the subsequent analysis; these are shown in Table~\ref{results}. Spectra from this fit for two of the observations (those with the highest and lowest total fluxes) are shown in Figure~\ref{fig:spectra}. The other fits in Table~\ref{spectral_param} explore other values for the NS parameters and tied power-law index. Allowing the distance to vary (fit 4) for fixed canonical values of the NS mass and radius yields a best-fit distance of $7.3\pm1.4$ kpc, consistent with the distance estimate from \citet{lin2009b}. Allowing instead the radius to vary (fit 5) gives $R_\mathrm{ns}=11.7\pm2.0$ km. We note that \citet{lin2009b} independently derive an asymptotic (apparent blackbody) radius of $8\pm1$ km from the three type I bursts and, correcting for the effects of redshift and spectral hardening (assuming a distance of 8.8 kpc, a NS mass of 1.4 $M_\sun$, and a hardening factor of 1.4), quote an actual NS radius of $\simeq$13 km. This value must, however, be regarded as highly uncertain, mainly due to the large uncertainty in the appropriate value for the hardening factor. Allowing the mass to vary (fit 6) gives $M_\mathrm{ns}=1.6\pm0.4\textrm{ }M_\sun$. Allowing both the radius and mass to vary (fit 7) gives a radius of $11.6\pm2.6$ km and a mass of $1.5\pm0.6\textrm{ }M_\sun$. The derived absorption column and power-law indices are largely unaffected by the values of the other parameters. The values of the absorption column and the power-law index for XMM-3 are tightly constrained to be $N_\mathrm{H}\simeq($1.9--2.0$)\times10^{22}\textrm{ cm}^{-2}$ and $\alpha'\simeq1.3$--1.4. This value for $N_\mathrm{H}$ is in good agreement with the value of $\simeq$$2.0\times10^{22}\textrm{ cm}^{-2}$ derived from \xte\ and \sw\ observations during the outburst \citep{lin2009b,lin2009a}. The combined (in some sense averaged) power-law index for the other observations (i.e., all the quiescent observations except XMM-3) is a less meaningful quantity, but seems to be higher than the XMM-3 index, having a value of $\alpha\simeq1.9\pm0.3$. The main fit has a reduced $\chi^2$ value, $\chi^2_\nu$, of 1.12 for 517 degrees of freedom (dof), corresponding to a $\chi^2$ probability, $P_\chi$, of 0.035. This is only a marginally adequate fit; possible reasons for this will be mentioned in Section~\ref{sec:tests}. The other fits in Table~\ref{spectral_param} all have $\chi^2_\nu$ in the range 1.11--1.13 for 515--518 dof.

From our main spectral fit we derive both unabsorbed and absorbed total fluxes, as well as unabsorbed fluxes for each of the two spectral components. We use the same fractional errors for the unabsorbed total fluxes as for the absorbed ones, thereby ignoring some error arising from uncertainty in the effects of the absorption on the flux values. However, given how tight our constraints on the absorption column are, this should be a relatively small effect. Since the temperature is the only free parameter in the {\tt nsatmos} model, we can simply propagate the error in the temperature to get the error in the unabsorbed thermal flux. The value of the normalization parameter of the {\tt pegpwrlw} model is the unabsorbed flux of that component (in a chosen energy range), and the error in that parameter therefore directly gives the error in the unabsorbed power-law flux.

\subsection{Cooling Curves}\label{sec:cooling_curves}

Figure~\ref{fig:rxte} shows the transition from the final stage of outburst to quiescence. Plotted is the total unabsorbed luminosity in the 0.5--10 keV band for the 37 \xte\ observations made in the period 2007 July 17--August 7, and the three \sw\ observations discussed above, as well as the first three \cxo\ and \xmm\ observations. The luminosity decreased by a factor of $\sim$2000 in the final $\simeq$13 days of the outburst before starting a much slower decay. This period of low-level and slowly changing (compared to the outburst phase) emission, taking place after the steep drop in luminosity, is what we refer to as the quiescent phase (see also the top panel in Figure~\ref{fig:lum+cooling+pl}). Low-level accretion may be occurring during quiescence, but this current phase is clearly distinct from the much more luminous and variable outburst phase, during which accretion took place at much higher rates (and which we also refer to as the period of ``active'' accretion). The end of the outburst is tightly constrained to have occurred sometime in the $\simeq$4.3 day interval between the final \sw\ observation and the first \cxo\ observation. To get a more precise estimate for the end time of the outburst, here denoted by $t_0$, we fit simple exponential decay curves through the three \sw\ data points and the three \cxo\ and \xmm\ points in Figure~\ref{fig:rxte}. From the intersection of those two curves we define $t_0$ as MJD 54322.13 (2007 August 10 03:06 UT), i.e., $\simeq$2.8 days before the first \cxo\ observation.

\begin{deluxetable*}{cccccccccccccc}
\tablecaption{Best-fit Cooling Curve Parameters for Temperature Data\label{cooling_param}}
\tablehead{\multicolumn{4}{c}{Spectral Fit Parameters} & \colhead{} & \multicolumn{4}{c}{Exponential Decay Fit} & \colhead{} & \multicolumn{4}{c}{Broken Power-law Fit} \\
\cline{1-4} \cline{6-9} \cline{11-14}
\colhead{$D$} & \colhead{$R_\mathrm{ns}$} & \colhead{$M_\mathrm{ns}$} & \colhead{$\alpha$\tablenotemark{a}} & \colhead{} & \colhead{$\tau$\tablenotemark{b}}  & \colhead{$kT_\mathrm{eq}$\tablenotemark{c}} & \colhead{$kT'$\tablenotemark{d}} & \colhead{$\chi^2_\nu$\tablenotemark{e}} & \colhead{} & \colhead{$\gamma_1$\tablenotemark{f}} & \colhead{$\gamma_2$\tablenotemark{g}} & \colhead{$t_\mathrm{b}$\tablenotemark{h}} &  \colhead{$\chi^2_\nu$\tablenotemark{i}}\\
\colhead{(kpc)} & \colhead{(km)} & \colhead{($M_\sun$)} & \colhead{} & \colhead{} & \colhead{(days)}  & \colhead{(eV)} & \colhead{(eV)} & \colhead{} & \colhead{} & \colhead{} & \colhead{} & \colhead{(days)} &  \colhead{}}
\startdata
(8.8) & (10) & (1.4) & 1.93 && $117_{-19}^{+26}$ & $125.0\pm0.9$ & $37.4\pm1.8$ & 0.84 && $0.027\pm0.013$ & $0.070\pm0.004$ & $33_{-10}^{+23}$ & 1.08\\
(7.5) & (10) & (1.4) & 1.93 && $118_{-19}^{+27}$ & $118.4\pm0.9$ & $35.2\pm1.7$ & 0.76 && $0.025\pm0.013$ & $0.070\pm0.004$ & $32_{-9}^{+21}$ & 1.04\\
(10.1) & (10) & (1.4) & 1.93 && $115_{-18}^{+25}$ & $131.0\pm0.9$ & $39.5\pm1.8$ & 0.97 && $0.030\pm0.013$ & $0.070\pm0.004$ & $34_{-11}^{+26}$ & 1.15\\
(8.8) & (11.6) & (1.5) & 1.93 && $117_{-19}^{+27}$ & $120.3\pm0.9$ & $35.9\pm1.7$ & 0.76 && $0.025\pm0.013$ & $0.070\pm0.004$ & $32_{-9}^{+21}$ & 1.04\\
(8.8) & (10) & (1.4) & (1.3) && $132_{-20}^{+27}$ & $128.2\pm0.7$ & $34.1\pm1.3$ & 2.89 && $0.030\pm0.011$ & $0.063\pm0.003$ & $35_{-11}^{+27}$ & 2.46\\
(8.8) & (10) & (1.4) & (2.5) && $111_{-17}^{+24}$ & $120.8\pm0.9$ & $41.5\pm2.1$ &1.75 && $0.023\pm0.017$ & $0.079\pm0.005$ & $30_{-8}^{+17}$ & 2.61
\enddata
\tablecomments{In each case, the data points used for the cooling curve fit were derived from a simultaneous spectral fit to all observations. Selected spectral fit parameters are shown in the first four columns; numbers in parentheses indicate parameters fixed during the spectral fitting. Errors quoted are at the $1\sigma$ Gaussian (68.3\%) confidence level. --- $^\mathrm{a}\textrm{ }$Combined power-law index for all observations except the third \xmm\ observation (XMM-3). $^\mathrm{b}\textrm{ }$Best-fit $e$-folding time of the decay. $^\mathrm{c}\textrm{ }$Best-fit constant offset to the decay. $^\mathrm{d}\textrm{ }$Best-fit normalization coefficient of the decay. $^\mathrm{e}\textrm{ }$Reduced $\chi^2$ for the fit, which had 8 dof in each case. $^\mathrm{f}\textrm{ }$Best-fit pre-break power-law slope. $^\mathrm{g}\textrm{ }$Best-fit post-break power-law slope. $^\mathrm{h}\textrm{ }$Best-fit break time between power laws. $^\mathrm{i}\textrm{ }$Reduced $\chi^2$ for the fit, which had 7 dof in each case.}
\end{deluxetable*}

Table~\ref{results} lists temperatures and fluxes derived from the main fit to the \cxo\ and \xmm\ spectra discussed in Section~\ref{sec:fitting_results}. Figure~\ref{fig:lum+cooling+pl} shows a plot using results from this fit. The top two panels show the total unabsorbed 0.5--10 keV luminosity and the inferred effective NS surface temperature (as observed at infinity). The first five data points, taken in the first $\simeq$175 days of quiescence, show a fast drop in temperature. However, the sixth data point (XMM-3; at $\simeq$226 days) shows a large increase in both temperature and luminosity, and the following \cxo\ observation (CXO-4) also has a higher inferred temperature than before the increase. This is inconsistent with the monotonic decrease in temperature expected for a cooling NS crust. The last six \cxo\ observations all have temperatures similar to or slightly lower than the one immediately preceding XMM-3 (i.e., CXO-3). We assume that those are unaffected by whatever caused the ``flare-like'' behavior in the sixth and seventh observations, and when fitting cooling models to the data we exclude both XMM-3 and CXO-4 but include the subsequent observations (although some fits excluding only XMM-3 were also made; see below). We defer further discussion of the flare to the end of this section and Section~\ref{sec:non-thermal}.

\begin{figure}
\centerline{\includegraphics[width=8.5cm,trim=20 15 50 35,clip=true]{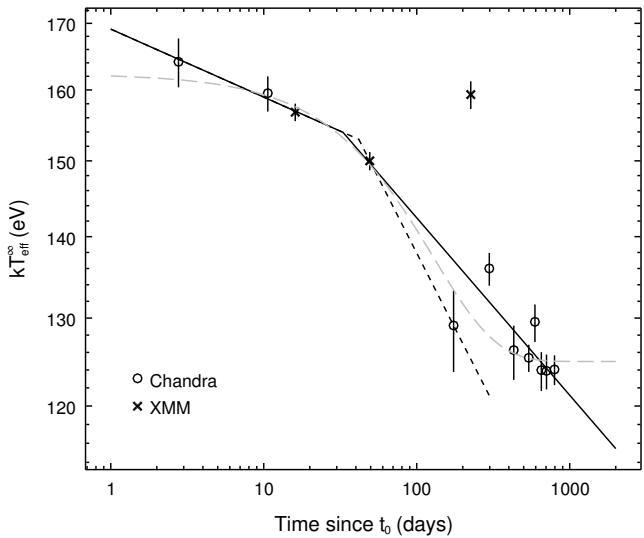}}
\caption{Effective NS surface temperature during quiescence, with best-fit cooling curves shown. The solid and short-dashed curves are broken power laws fitted to data points 1--5 plus 8--13, and data points 1--5, respectively. The gray long-dashed curve is the best-fit exponential decay curve with a constant offset (also shown in Fig.\ \ref{fig:lum+cooling+pl}).}\label{fig:cooling_log}
\end{figure}

We will now describe our fitting of the derived temperatures with cooling curve models. All the fits were performed with {\it Sherpa}, CIAO's modeling and fitting package \citep{freeman2001}; errors were estimated with the confidence method.\footnote{See documentation at the {\it Sherpa} Web site: \url{http://cxc.harvard.edu/sherpa/}.} We first fitted our temperature data with an exponential decay cooling curve plus a constant offset, i.e., a function of the form $T_\mathrm{eff}^\infty(t)=T'\exp[-(t-t_0)/\tau]+T_\mathrm{eq}$, with $t_0$ kept fixed at the value mentioned above. Shifts in the value of $t_0$ do not affect derived values for $\tau$ or $T_\mathrm{eq}$. The flare observations XMM-3 and CXO-4 were excluded from the fitting. We performed the temperature fit for data from the main spectral fit (1 in Table~\ref{spectral_param}), and also for spectral parameter values corresponding to five other fits (2, 3, 7, 8, and 9), to gauge the effects on the cooling fit parameters. The derived parameter values are shown in Table~\ref{cooling_param}. The main fit cooling curve is shown in Figure~\ref{fig:lum+cooling+pl} along with the best-fit constant offset (dashed line). The best-fit $e$-folding time is $\tau=117_{-19}^{+26}$ days with an offset of $T_\mathrm{eq}=125.0\pm0.9$ eV. For the other values of the NS parameters (mass, radius, and distance), the temperature values are systematically shifted by typically 5--10 eV, but the derived decay timescale is not affected to a significant extent. The effects of changing the value of the tied power-law index will be discussed in Section~\ref{sec:tests}. Including CXO-4 in the fit (but still excluding XMM-3) gives a longer timescale of $\tau=187_{-39}^{+49}$ days; the equilibrium temperature is not significantly affected.

\begin{deluxetable*}{cccccccccccccc}
\tablecaption{Best-fit Cooling Curve Parameters for Bolometric Thermal Luminosity Data\label{cooling_param_lum}}
\tablehead{\multicolumn{4}{c}{Spectral Fit Parameters} & \colhead{} & \multicolumn{4}{c}{Exponential Decay Fit} & \colhead{} & \multicolumn{4}{c}{Broken Power-law Fit}\\
\cline{1-4} \cline{6-9} \cline{11-14}
\colhead{$D$} & \colhead{$R_\mathrm{ns}$} & \colhead{$M_\mathrm{ns}$} & \colhead{$\alpha$\tablenotemark{a}} & \colhead{} & \colhead{$\tau$\tablenotemark{b}}  & \colhead{$L_\mathrm{eq}$\tablenotemark{c}} & \colhead{$L'$\tablenotemark{d}} & \colhead{$\chi^2_\nu$\tablenotemark{e}} & \colhead{} & \colhead{$\gamma_1$\tablenotemark{f}} & \colhead{$\gamma_2$\tablenotemark{g}} & \colhead{$t_\mathrm{b}$\tablenotemark{h}} &  \colhead{$\chi^2_\nu$\tablenotemark{i}}\\
\colhead{(kpc)} & \colhead{(km)} & \colhead{($M_\sun$)} & \colhead{} & \colhead{} & \colhead{(days)}  & \colhead{($10^{33} \textrm{ erg s}^{-1}$)} & \colhead{($10^{33} \textrm{ erg s}^{-1}$)} & \colhead{} & \colhead{} & \colhead{} & \colhead{} & \colhead{(days)} &  \colhead{}}
\startdata
(8.8) & (10) & (1.4) & 1.93 && $\phn88_{-14}^{+18}$ & $5.35\pm0.14$ & $\phn9.83\pm0.61$ & 0.80 && $0.108\pm0.053$ & $0.282\pm0.017$ & $33_{-10}^{+21}$ & 1.14\\
(7.5) & (10) & (1.4) & 1.93 && $\phn88_{-14}^{+19}$ & $5.94\pm0.16$ & $10.80\pm0.68$ & 0.71 && $0.100\pm0.053$ & $0.281\pm0.017$ & $32_{-9}^{+19}$ & 1.12\\
(10.1) & (10) & (1.4) & 1.93 && $\phn87_{-13}^{+18}$ & $4.90\pm0.12$ & $\phn9.10\pm0.56$ & 0.92 && $0.119\pm0.053$ & $0.283\pm0.016$ & $34_{-10}^{+23}$ & 1.18\\
(8.8) & (11.6) & (1.5) & 1.93 && $\phn88_{-14}^{+18}$ & $5.87\pm0.16$ & $10.71\pm0.67$ & 0.72 && $0.101\pm0.053$ & $0.281\pm0.017$ & $32_{-9}^{+19}$ & 1.13\\
(8.8) & (10) & (1.4) & (1.3) && $105_{-16}^{+21}$ & $5.92\pm0.11$ & $\phn9.15\pm0.50$ & 2.89 && $0.119\pm0.044$ & $0.255\pm0.013$ & $35_{-11}^{+25}$ & 2.37\\
(8.8) & (10) & (1.4) & (2.5) && $\phn75_{-10}^{+13}$ & $4.65\pm0.14$ & $10.73\pm0.69$ & 1.92 && $0.091\pm0.067$ & $0.316\pm0.020$ & $28_{-8}^{+15}$ & 3.76
\enddata
\tablecomments{All notes for Table~\ref{cooling_param} apply here as well.}
\end{deluxetable*}

As will be discussed in Section~\ref{sec:thermal}, a more physically motivated cooling curve model is a broken power law leveling off to a constant at late times. We therefore also fitted a broken power-law model, excluding XMM-3 and CXO-4 as before, to temperature data corresponding to the same six spectral fits as before. The derived break times and power-law slopes are shown in Table~\ref{cooling_param}. The best-fit broken power-law curve to data from the main spectral fit is shown in Figure~\ref{fig:cooling_log} (solid curve). The data indicate that a break in the model is needed; a simple power law does not provide an adequate fit ($\chi^2_\nu=2.45$ for 9 dof, compared to $\chi^2_\nu=1.08$ for 7 dof in the broken power-law case). The best-fit break time is $t_\mathrm{b}=33_{-10}^{+23}$ days post-outburst; the best-fit slopes are $\gamma_1=0.027\pm0.013$ (pre-break) and $\gamma_2=0.070\pm0.004$ (post-break). These values do not change significantly when the other values of the spectral parameters are used (see Table~\ref{cooling_param}). The exponential decay and broken power-law fits have $\chi^2_\nu\simeq$1 when the combined power-law index is free to vary in the spectral fit; the values for the exponential fits tend to be slightly lower. For the spectral fits with the power-law index fixed, the $\chi^2_\nu$ values are considerably higher ($\simeq$2--3); this is in large part due to the smaller error bars for the temperatures resulting from the fixing of the index. Including CXO-4 does not significantly affect the broken power-law fit.

Given that the broken power-law cooling curve is expected to level off at some point, and that from CXO-3 onward the observations only show small changes in temperature (not counting the flare observations), we also fitted a broken power law to only the first five data points from the main spectral fit (short-dashed curve in Figure~\ref{fig:cooling_log}). This gave the same pre-break slope as before, a break time $t_\mathrm{b}=41_{-9}^{+131}$ days, and a post-break slope $\gamma_2=0.119\pm0.030$. The upper limit to the break time is in this case only constrained by the time of the fifth observation. The time of the break is in general poorly constrained by our fits. In all cases, the $\chi^2$ surface in parameter space, although locally quadratic around the minimum, shows significant deviations from a parabolic shape toward higher break times, in some cases before a difference in the value of $\chi^2$ corresponding to a $1\sigma$ error is reached. Any upper limits to $t_\mathrm{b}$ given should therefore be regarded with caution. Overall, we conclude that the time of the break could be as early as $\simeq$20 days and as late as $\simeq$150 days post-outburst. Since a break time after XMM-2 (which is at $\simeq$49 days post-outburst) is clearly also allowed by the data, we fitted a simple power law to the first four data points to estimate what the pre-break slope would be in that case. This gave $\gamma_1=0.035\pm0.007$, slightly higher than for fits with a break before XMM-2. We note that, in contrast to the exponential decay fits, the results from the power-law fits depend (sensitively) on the value used for $t_0$. Placing $t_0$ 2.5 days later (i.e., $\simeq$0.3 days before the first observation in quiescence) and fitting the entire data set (except XMM-3 and CXO-4) gives $\gamma_1=0.012\pm0.006$ and $t_\mathrm{b}=26_{-6}^{+7}$ days. However, results from fits with $t_0$ very close to the first data point should be regarded with caution. Due to the divergence of the power-law model at $t_0$ (since $T_\mathrm{eff}^\infty(t)\propto(t-t_0)^{-\gamma_1}$), the power-law behavior of the cooling curve cannot extend all the way back to $t_0$; the curve has to flatten out from a power law before that. We also performed a fit with $t_0$ placed 13 days earlier, i.e., around the time when the decay of the outburst light curve steepened dramatically (see Figure~\ref{fig:rxte}). Although the source was still in outburst at that time, it may be appropriate to consider the effective start time of the cooling as the onset of the fast decrease in accretion rate implied by the steep decay. In this case, there is no indication of a break in the cooling curve; the entire curve is well fitted with a single power law with slope $0.073\pm0.003$, even when including CXO-4.

\begin{figure}
\centerline{\includegraphics[width=8.5cm,trim=15 17 40 30,clip=true]{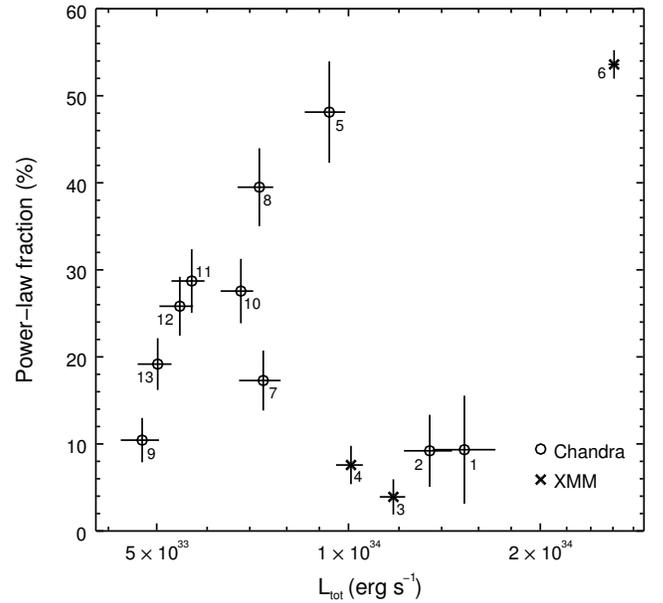}}
\caption{Fractional contribution of the power-law component to the total unabsorbed flux in the 0.5--10 keV band as a function of total unabsorbed luminosity in the same energy range. The numbers indicate the time sequence of the observations.}\label{fig:pl_frac}
\end{figure}

In addition to the fits to the temperature data discussed above, we also made analogous fits to the unabsorbed bolometric luminosity of the thermal component. This was done to permit possible comparison with results from other sources, and because such fits are of some interest to theorists. The results are given in Table~\ref{cooling_param_lum} and will not be discussed further in this paper, apart from noting that the extrapolation of the {\tt nsatmos} model outside the energy range covered by our data gives rise to some systematic uncertainty. Our observed 0.5--10 keV thermal flux represents $\simeq$80\%--85\% of the inferred bolometric thermal flux (practically all of which is contained in the 0.01--10 keV band). We therefore expect the induced systematic error in the bolometric thermal flux due this extrapolation to be at most $\simeq$10\%.

The non-thermal flux has varied irregularly throughout the quiescent phase (see Table~\ref{results} and the bottom panel of Figure~\ref{fig:lum+cooling+pl}). As mentioned before, XMM-3 has by far the largest non-thermal flux, $\simeq$$1.5\times10^{-12}\textrm{ erg s}^{-1}\textrm{ cm}^{-2}$ in the 0.5--10 keV band, corresponding to an unabsorbed luminosity of $\simeq$$1.4\times10^{34}\textrm{ erg s}^{-1}$ for a distance of 8.8 kpc. Values for the other observations range from $\simeq$$5\times10^{-14}\textrm{ erg s}^{-1}\textrm{ cm}^{-2}$ to $\simeq$$5\times10^{-13}\textrm{ erg s}^{-1}\textrm{ cm}^{-2}$. A plot of the fractional contribution of the non-thermal flux to the total unabsorbed 0.5--10 keV flux as a function of the total unabsorbed 0.5--10 keV luminosity is shown in Figure~\ref{fig:pl_frac}. We note that changes in the NS parameters in the spectral fit yield systematic shifts of $\simeq$(1--4)$\times10^{-14}\textrm{ erg s}^{-1}\textrm{ cm}^{-2}$ in the derived power-law fluxes. The fractional changes are in the range $\simeq$7\%--40\% (except for XMM-3 where the changes are $\simeq$1\%--2\%). The effects of changing the value of the tied power-law index will be discussed in Section~\ref{sec:tests}.

\subsubsection{Further Tests of Cooling Results}\label{sec:tests}

Since we use data from both \cxo\ and \xmm, cross-calibration error between the instruments can possibly affect our results, especially given how important the second \xmm\ observation is in determining the timescale of the decay. The cross-calibration error between \cxo\ ACIS and \xmm\ EPIC fluxes is in general expected to be $\lesssim$10\% (H. Marshall 2008, private communication; \citealt{snowden2002,stuhlinger2008}). Given the fourth power temperature dependence of the bolometric thermal flux, the relative cross-calibration error in the temperatures is likely at most a few percent. To estimate the effect on our results of a systematic shift in the \xmm\ temperatures with respect to the \cxo\ ones, we increased/decreased the \xmm\ temperatures by  $\pm3\%$ and repeated the fits. For the exponential decay fit, this yielded $e$-folding times of $\tau=149_{-24}^{+34}$ days and $\tau=87_{-13}^{+18}$ days, respectively, and no significant change in the equilibrium temperature. In both cases the quality of the fit was worse than without any shifting. The shifts had a greater impact on the broken power-law fits. A +3\% shift gave $\gamma_1=0.007\pm0.013$, $\gamma_2=0.082\pm0.004$, and $t_\mathrm{b}=30_{-6}^{+9}$ days, with a decrease in the quality of the fit. After a --3\% shift there is no longer need for a break in the curve; the entire cooling curve is well fitted with a single power law with slope $0.054\pm0.002$ (still excluding XMM-3 and CXO-4 from the fit, although CXO-4 is in reasonably good agreement with the best-fit curve, and including it does not change the estimate of the slope).

To get an indication of how sensitive our results are to our choice of a model for the non-thermal component, we repeated our main spectral fit with the power-law model replaced by the {\tt simpl} model in XSPEC. This is an empirical convolution model for Comptonization, which converts a fraction of input seed photons to a power law \citep{steiner2009}. The model has only two free parameters, the power-law index and the fraction of scattered photons, and can be used with any spectrum of seed photons. Compared to our previous results, we see small temperature shifts in the range $-0.8$ to 1.5 eV for all but three of the observations; for XMM-3, CXO-3, and CXO-5 (the three observations with the largest non-thermal components) we get larger shifts of 14.7, 5.5, and 3.5 eV, respectively. The values of the absorption column and power-law indices are similar to before, as are the $\chi^2_\nu$ values. The behavior of the non-thermal component also seems to be similar, although comparison is complicated by the fact that when using {\tt simpl} the thermal and non-thermal components cannot be disentangled in the same manner as when a regular power law is used. The derived $e$-folding time for these new temperature values is $\tau=128_{-21}^{+29}$ days with a best-fit constant offset of $126.0\pm0.8$ eV; these values are not significantly different from the previous ones. The quality of the exponential cooling curve fit is significantly worse than before, with $\chi^2_\nu=1.74$ for 8 dof. In the case of the broken power-law fit, using {\tt simpl} instead of a power-law model does not significantly change the derived parameter values, but does adversely affect the quality of the fit.

\begin{figure}
\centerline{\includegraphics[width=6cm,trim=15 55 35 60,clip=true]{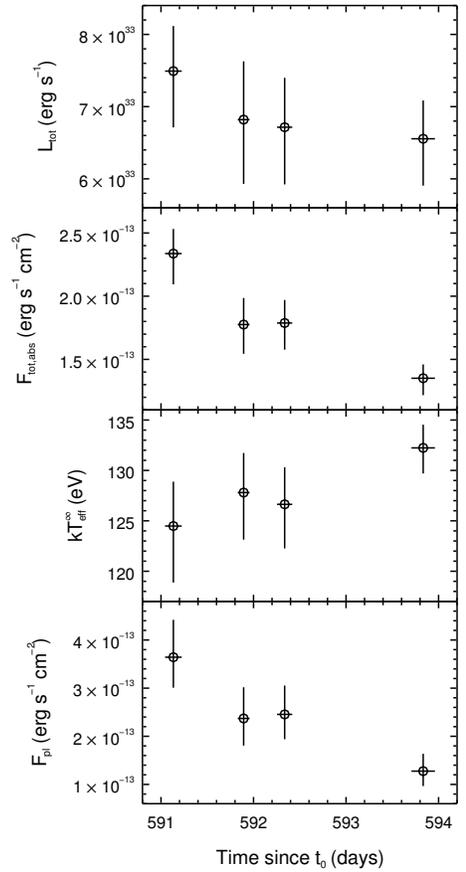}}
\caption{Total unabsorbed luminosity in the 0.5--10 keV band (top/first panel), total absorbed flux in the 0.5--10 keV band (second panel), effective NS surface temperature (third panel), and unabsorbed power-law flux in the 0.5--10 keV band (bottom/fourth panel) for the four sub-exposures in the seventh {\it Chandra} observation. The horizontal error bars indicate the durations of the exposures.}\label{fig:cxo-7}
\end{figure}

As mentioned in Section~\ref{sec:fitting_results}, the seventh \cxo\ observation consisted of four separate exposures taken over approximately 3 days, and the tenth \cxo\ observation consisted of two exposures taken over a period of $\simeq$27 hr. To look for possible variability between exposures within these two observations, we performed a spectral fit to all the \cxo\ and \xmm\ observations, identical to the main fit described in Section~\ref{sec:fitting_results} apart from the fact that the temperature and power-law normalization parameters were not tied between the four CXO-7 spectra and the two CXO-10 spectra. Untying these parameters within CXO-7 and CXO-10 led to a considerable improvement of the overall fit, from $\chi^2_\nu=1.12$ for 517 dof to $\chi^2_\nu=1.07$ for 509 dof ($P_\chi=0.14$). Figure~\ref{fig:cxo-7} shows a plot of the temperature and various fluxes for the four exposures in CXO-7. Although not shown in the plots, we note that the count rate exhibits a decrease similar to that seen for the absorbed flux; the net count rates in the first and last exposures (in the 0.5--10 keV band) are $(1.91\pm0.12)\times10^{-2}$ and $(1.57\pm0.09)\times10^{-2}\textrm{ counts s}^{-1}$, respectively. The plot indicates a possible decrease in the power-law flux over the 3 day period. It is not implausible that this observed decrease is real, especially if the non-thermal component arises from accretion. However, the fact that the behavior of the power-law flux seems to be anti-correlated with the behavior of the temperature (which shows a slight but, given the errors, non-significant increase) may point to a limitation in our ability to separate the contributions from the thermal and non-thermal components to the total flux. Looking at the CXO-10 exposures further indicates that this may be the case. The net count rates observed are $(1.26\pm0.06)\times10^{-2}$ and $(0.95\pm0.06)\times10^{-2}\textrm{ counts s}^{-1}$ in the first and second exposures. A short-term light curve suggests an overall decrease in net count rate during the first exposure; a one-sided Kolmogorov--Smirnov (K--S) test comparing photon arrival times to a uniform count rate model rejects the null hypothesis of a constant light curve at a 97.7\% level. However, in this case the spectral fits show no change in the power-law flux, but a decrease in temperature from $127.2\pm1.9$ to $119.5\pm2.6$ eV. It is clearly not plausible for the surface temperature to change in such a short time; any real change in flux is likely due to the non-thermal component (unless, perhaps, some of the thermal flux arises from accretion, see discussion in Section~\ref{sec:non-thermal}). Indeed, constraining the two temperatures to be the same and only allowing the power-law normalization to vary independently results in a perfectly adequate fit to the two spectra; this is also the case for the four spectra in CXO-7. It is therefore likely that any observed differences in temperatures within CXO-7 and CXO-10 are simply an artifact of the spectral fitting, and it is possible that similar effects are in general affecting our derived temperatures and fluxes to some extent. We note that for the overall fit where the temperatures were tied within CXO-7 and CXO-10, but the power-law normalizations allowed to vary, the derived temperatures are $129.0\pm2.2$ eV for CXO-7 and $124.3\pm1.6$ eV for CXO-10; these values are not significantly different from the ones in the main spectral fit. In this case, the 0.5--10 keV unabsorbed power-law flux for the first and last exposures in CXO-7 is $(3.2\pm0.5)\times10^{-13}$ and $(1.5\pm0.3)\times10^{-13}\textrm{ erg s}^{-1}\textrm{ cm}^{-2}$; for the two CXO-10 exposures, the power-law fluxes are $(1.2\pm0.2)\times10^{-13}$ and $(0.7\pm0.2)\times10^{-13}\textrm{ erg s}^{-1}\textrm{ cm}^{-2}$.

Another possible cause for concern is that tying the power-law index between the different observations (except for XMM-3) may skew the derived values for the temperature and non-thermal flux. To gauge to what extent our results may be affected by this, we repeated our spectral fit twice, with the index fixed at values of 1.3 (similar to the value for XMM-3) and 2.5. Fixing the index at 1.3 resulted in all the temperature values increasing. All but two had shifts in the range 0.9--3.8 eV; for CXO-3 and CXO-5 the shifts were 8.1 eV and 5.6 eV, respectively. The (non-XMM-3) power-law fluxes all decreased by 24\%--34\%. Fixing the index at 2.5 resulted in temperature shifts in the range $-5.8$ to $-0.1$ eV, except for CXO-3 and CXO-5, whose shifts were $-16.2$~eV and $-9.1$ eV. The power-law fluxes increased by 59\%--78\%. The derived parameters for the cooling curve fits in both these cases are shown in Table~\ref{cooling_param}; no drastic changes are seen. Overall, for most of the observations it is unlikely that the individual temperatures are skewed by more than $\simeq$6 eV due to the tying of the power-law index. For CXO-3 and CXO-5, the two observations with the largest non-thermal components (excluding XMM-3), the effect may be larger. However, we note that when the indices for these two observations are allowed to vary independently from the others, the temperatures are only shifted by $-5$~eV (CXO-3) and +1.6 eV (CXO-5). The values of the two indices in this case are $2.2\pm0.4$ and $1.8\pm0.4$. The power-law fluxes are likely skewed by less than a factor of 2 in all cases. While these possible effects on the temperatures and fluxes are unfortunate, the situation would probably not be improved by allowing the power-law indices to vary independently. In that case the indices take on widely varying (and in some cases unphysically high or low) values, many of which are very poorly constrained; almost all the $1\sigma$ confidence intervals for the free index values overlap. In addition, the overall fit becomes very unstable. Given all this, it is unlikely that the temperatures and fluxes derived with the free indices are in general any more accurate than those derived using a single ``average'' value for the index, and for those observations where the indices take on extreme values they are probably less accurate (although this could be mitigated to some extent by constraining the index values to lie in a certain physically plausible range, e.g., 1--2.5). We conclude that the best solution, although clearly not perfect, is to tie the index.

It is natural to ask whether this possible skewing of the temperatures, and/or limitations in our ability to separate the contributions of the two spectral components to the total flux, can explain the anomalously high temperatures seen for XMM-3 and CXO-4. To test this we fitted the spectra from these observations individually, fixing the temperatures at the value of the best-fit exponential decay curve to the main spectral fit (see Figure~\ref{fig:lum+cooling+pl}), and allowing the power-law indices to vary. The absorption column was fixed at the best-fit value from the main spectral fit. The CXO-4 spectrum is well fitted in this manner ($\chi^2_\nu=0.80$ for 15 dof), although the value of the power-law index is high, $2.96\pm0.25$ (and very different from the value attained when the temperature is also allowed to vary, $0.7\pm1.1$, in which case the temperature increases from the original main spectral fit value). Constraining the temperature at the value of the best-fit broken power-law curve (see Figure~\ref{fig:cooling_log}) results in a lower value for the index, $2.55\pm0.33$ (and still giving an excellent fit). In light of this, and the fact that CXO-4 does not show an anomalously high luminosity or non-thermal flux compared to the other observations (in contrast to XMM-3), we conclude that it is quite possible that the high temperature derived from the main spectral fit is in this case simply an artifact of the fitting process, and not due to a real increase in the temperature compared to CXO-3. Fixing the temperature in XMM-3 to the values of the exponential decay or broken power-law curves gives unacceptable fits; the $\chi^2_\nu$ values are 1.51 and 1.46 for 225 dof ($P_\chi$ of  $1\times10^{-6}$ and $9\times10^{-6}$). In comparison, the $\chi^2_\nu$ value is 1.18 ($P_\chi=0.033$) for the original XMM-3 fit with the temperature free; much better, but admittedly only marginally acceptable. This is partly due to the fact that the model underestimates the lowest-energy part of the spectrum; this can be seen in the pn spectrum in Figure~\ref{fig:spectra}. Indeed, allowing the absorption column to vary leads to a considerably better fit with $\chi^2_\nu=1.11$ ($P_\chi=0.13$). In this case $N_\mathrm{H}=(1.72\pm0.05)\times10^{22}\textrm{ cm}^{-2}$; the derived temperature is $150.2\pm3.2$ eV. Untying the absorption column for XMM-3 from that of the other observations leads to a significant improvement in the overall main fit, from a $\chi^2_\nu$ value of 1.12 for 517 dof ($P_\chi=0.035$) to 1.07 ($P_\chi=0.12$); the tied $N_\mathrm{H}$ then assumes a value of $1.99\pm0.03\textrm{ cm}^{-2}$. Furthermore, doing this in conjunction with untying the temperatures and power-law normalizations within CXO-7 and CXO-10, as discussed earlier in this section, leads to a very good fit with $\chi^2_\nu=1.02$ for 508 dof ($P_\chi=0.34$). We also note that the individual XMM-3 fit, with $N_\mathrm{H}$ free and the temperature fixed to the value of the best-fit broken power-law curve, gives $\chi^2_\nu=1.17$ ($P_\chi=0.042$); a marginally acceptable fit. However, the much lower value of the absorption column for XMM-3 (in this case $1.59\pm0.04\textrm{ cm}^{-2}$), compared to the tied value for the rest of the observations, is rather implausible. Overall, we conclude that the XMM-3 spectra are hardly compatible with a temperature in line with the overall trend of the rest of the observations, and that there is in this case likely a real and significant increase in thermal flux compared to the previous observation. The high luminosity and non-thermal flux also unequivocally show that this observation is quite distinct in behavior from the other ones.

\section{Discussion}\label{sec:discussion}

We have presented \xte\ and \sw\ observations tracking the final 3 weeks of the 2006--2007 outburst of XTE J1701--462, and subsequent \cxo\ and \xmm\ observations monitoring the source during the first $\simeq$800 days of quiescence. The transition from active accretion to quiescence is resolved with much better precision than for any other cooling NS transient observed after an extended outburst; the end of the outburst is tightly constrained to a $\simeq$4 day window in 2007 August. We fit the spectra obtained during quiescence with a two-component model consisting of a NS atmosphere model (thermal component) and a power-law model (non-thermal component). The effective surface temperature of the NS, derived from the thermal component, was seen to decay rapidly in the first $\sim$200 days of quiescence, which we interpret as the cooling of the NS crust toward thermal equilibrium with the core, after having been heated by accretion during the outburst. Our data set yields a much better sampled cooling curve than those of other cooling NS transients observed to date. The interpretation of the data is complicated by an apparent temporary increase in the temperature $\simeq$220 days into quiescence. The existence of the non-thermal component in the quiescent spectra also adversely affects how well we can constrain the behavior of the thermal component. The non-thermal flux from the source has varied irregularly throughout the quiescent phase by a factor of $\simeq$30, representing $\simeq$5\%--50\% of the total flux. Fitting the inferred temperatures (excluding the two observations showing the temporary increase) with an exponential decay plus a constant offset yields an $e$-folding time of $\simeq$$120^{+30}_{-20}$ days. This value is not affected by uncertainties in the NS distance, radius, or mass, and uncertainty in the modeling of the non-thermal component likely has only a small effect on the derived timescale. The exponential fit implies that the temperature has reached a roughly constant value of $\simeq$125 eV (assuming the best-estimate distance of 8.8 kpc). Allowing for uncertainty in the NS distance, mass, and radius, and in the modeling of the non-thermal component, gives a possible error of $\simeq$15 eV. This best-fit baseline temperature corresponds to a bolometric thermal luminosity (redshifted and unabsorbed) of $\simeq$$5.4\times10^{33}\textrm{ erg s}^{-1}$. The temperature data can also be adequately fitted with a broken power law (but not a simple power law). Taking into account that the cooling curve may have flattened out from the power-law behavior, the break time is constrained to $\simeq$20--150 days after the outburst end.

\subsection{Transition to Quiescence}\label{sec:transition}

\sj\ was observed to transition sharply from active accretion to quiescence in the final 2 weeks of the outburst. The decay rate increased dramatically around 2007 July 29 and the luminosity subsequently dropped by a factor of $\sim$2000 in the final $\simeq$13 days before entering quiescence (see Figure~\ref{fig:rxte}). The drop is well represented by a simple exponential decay with an $e$-folding time of $\simeq$1.7 days. This evolution from outburst to quiescence is similar to that seen in some other NS X-ray transients, e.g., Aql X-1. \citet{campana1998b} analyze observations of Aql X-1 at the end of a 1997 outburst in which the decay steepened suddenly and the luminosity then decreased by 3 orders of magnitude in less than 10 days; the decrease is well described by an exponential decay with an $e$-folding time of $\simeq$1.2 days. This behavior, closely resembling that seen for \sj, was interpreted by \citet{campana1998b} as being caused by the onset of the propeller mechanism, which impedes accretion, and signaling the turning on of a rotation-powered pulsar. They interpret the subsequent quiescent emission as arising from a shock between the pulsar wind and outflowing material from the companion star. However, several serious problems with this general interpretation of the steepening of the outburst decay rate in NS transients have been pointed out \citep[see, e.g.,][]{jonker2004}. These are both observational in nature, such as the fact that steepening of the decay has also been seen in black hole transients \citep[e.g.,][]{jonker2004b}, as well as theoretical \citep[e.g.,][]{rappaport2004}.

\subsection{Behavior of the Thermal Component}\label{sec:thermal}

It is useful to compare the behavior of \sj\ to that of the cooling quasi-persistent transients KS 1731--260 and MXB 1659--29. Following their $\simeq$12.5 and $\simeq$2.5 yr outbursts, the effective surface temperatures of KS 1731--260 and MXB 1659--29 were seen to decay exponentially with $e$-folding times of $305\pm47$ and $465\pm25$ days, respectively, reaching an approximately constant level in $\sim$1000--1500 days \citep{wijnands2001,wijnands2002a,wijnands2002b,rutledge2002,wijnands2003,wijnands2004a,wijnands2004b,cackett2006,cackett2008}. However, as mentioned in Section~\ref{sec:intro}, a recent observation of KS 1731--260 indicates that the source may still be cooling slowly. It was pointed out early on \citep{wijnands2002a,wijnands2004a} that the KS 1731--260 and MXB 1659--29 data indicated  NS crusts with high thermal conductivity. More recently, \citet{shternin2007} compared simulations of deep crustal heating and subsequent crustal cooling to the observations of KS 1731--260, and reached the conclusion that low thermal conductivity, corresponding to an amorphous (i.e., non-crystalline) crust, is inconsistent with the data. This is in agreement with the results of molecular dynamics simulations \citep{horowitz2007, horowitz2009a, horowitz2009b}, which indicate that the crust of an accreting NS will form an ordered crystal. \citet[][hereafter BC09]{brown2009} construct models of the thermal relaxation of a NS crust following an extended accretion episode and confirm the finding of \citet{shternin2007} that the thermal conductivity of the crust is high. Fitting their models to the observations of KS 1731--260 and MXB 1659--29, BC09 are able to place constraints on the crust parameters of the NSs; in particular, they are able to tightly constrain the so-called impurity parameter (which measures the distribution of the nuclide charge numbers, $Z$, in the crust material) for MXB 1659--29. They find a low value for the parameter, indicating a small amount of impurities in the crust (i.e., material with a small spread in $Z$) and high conductivity. In the case of \sj, the fact that the source seems to have cooled considerably faster than both KS 1731--260 and MXB 1659--29 (the derived exponential decay timescale is shorter by factors of $\simeq$2--5) strongly indicates a highly conductive crust, and possibly suggests low-impurity material. Extracting the exact implications of our data requires fitting of theoretical models; however, that is beyond the scope of this paper.

The models of BC09 indicate that the temperature cooling curve should actually not follow an exponential decay, but should rather be close to a broken power law flattening out to a constant at late times. The break is predicted to take place a few hundred days after the end of the outburst, and is set by the thermal diffusion time to the surface from the depth in the crust at which the material transitions from a classical to a quantum crystal, close to neutron drip (which is where neutrons start to leak out of nuclei, thereby producing a neutron gas between the nuclei). The break is mainly due to the suppression of the specific heat of the crust material with increasing density. Establishing whether and when a break in an observed cooling curve takes place would provide valuable input for theoretical models of the NS interior. There is no indication of a break having occurred in the cooling curve of KS 1731--260; the temperature data are actually well fitted with a single power law of slope $0.12\pm0.01$ \citep{cackett2008}. A break may have occurred for MXB 1659--29 in the first $\sim$500 days after the end of the outburst (the fits of BC09 imply a break at $\simeq$300--400 days post-outburst), but the scarcity of observations precludes drawing firm conclusions. As can be seen in Figure~\ref{fig:cooling_log}, there is a strong indication for a break in the cooling curve of \sj\ (although see caveats mentioned in Section~\ref{sec:cooling_curves}). The time of the possible break, however, is only $\simeq$20--150 days post-outburst, much earlier than the break predicted by BC09. It is therefore not clear whether the break seen for \sj, if real, is the one predicted by BC09, or is perhaps an unrelated extra structure in the cooling curve. Judging from Figure~10 in BC09, the time of the break can perhaps be decreased to as little as $\sim$150 days by assuming a rock-bottom value of 0 for the impurity parameter (no impurities in the crust), corresponding to a very high thermal conductivity. In addition, BC09 point out that the timescale of the cooling is proportional to $R_\mathrm{ns}^4M_\mathrm{ns}^{-2}(1+z)^{-1}$. Their model assumes $R_\mathrm{ns}=11.2$ km and $M_\mathrm{ns}=1.62\textrm{ M}_\sun$; a smaller radius and/or a larger mass (although a large mass is somewhat unlikely for the NS in \sj; see discussion below) would therefore push the break to an earlier time. This suggests that the observed break could conceivably be the one predicted by BC09 if it is in the upper part of the $\simeq$20--150 day range. A break time in the lower part of the range is not consistent with the predicted break. A possible alternative explanation for the break is the existence of a strong nuclear heat source in the outer crust. Such a source would alter the shape of the crust temperature profile in surrounding layers during outburst. This could cause a break in the cooling curve around a time corresponding to the thermal diffusion time from the depth of the source to the surface. Interestingly, \citet{horowitz2008} calculate that ${}^{24}$O should fuse at densities near $10^{11}\textrm{ g cm}^{-3}$, releasing 0.52 MeV per accreted nucleon. According to the models of BC09, the thermal diffusion time to the surface from a depth corresponding to a density of $10^{11}\textrm{ g cm}^{-3}$ is $\sim$100 days (see Figures 5 and 6 in their paper); this is consistent with the time of the break in the J1701 cooling curve.

In the models of BC09, the initial slope of the broken power-law curve gives a direct measure of the inward flux near the top of the crust during outburst. They note that observations in the first 2 weeks after the end of an outburst are critical for constraining the depth and strength of heat sources in the outermost layers of the crust. \sj\ was observed 3 times in the first $\simeq$16 days of quiescence; these observations should be able to provide valuable input for theoretical models. We note that our data set is unique in this respect. MXB 1659--29 and KS 1731--260 were not observed until at least $\simeq$31 and $\simeq$48 days, respectively, had passed in quiescence \citep{cackett2006}; the situation is less clear with EXO 0748--676, since the end time of the outburst is only constrained to a $\simeq$7 week window \citep{degenaar2009}. Our data, in conjunction with the results of BC09, allow us to calculate an estimate for the energy release per accreted nucleon in the outermost layers of the crust. Using Equation~12 in BC09 and our estimate of $\simeq$0.027 for the pre-break slope of the cooling curve gives an outer crust flux during outburst of $\simeq$$9.0\times10^{20}\textrm{ erg s}^{-1}\textrm{ cm}^{-2}$ and, integrating over a surface of radius 10 km, a total energy flow rate of $\simeq$$1.1\times10^{34}\textrm{ erg s}^{-1}$. The observed bolometric energy output of the outburst, $\simeq$$1.0\times10^{46}$ erg (see Section~\ref{sec:1701}), gives an average luminosity of $\simeq$$2.0\times10^{38}\textrm{ erg s}^{-1}$ over the approximately 19-month-long outburst. Assuming an accretion-powered luminosity, $L=\epsilon\dot{M}c^2$ with $\epsilon=0.2$, we get an estimate of $\langle\dot{M}\rangle\simeq1.1\times10^{18}\textrm{ g s}^{-1}\simeq1.7\times10^{-8}\textrm{ }M_\sun\textrm{ yr}^{-1}$ for the average mass accretion rate during the outburst. The total energy flow rate derived from the early-time slope then corresponds to $\simeq$11 keV per accreted nucleon. This value has a large uncertainty and is probably an underestimate, since the accretion rate was lower than the average value in the later parts of the outburst. Accounting for the inferred outer crust flux therefore requires more energy per accreted nucleon in the outermost layers of the crust; these are the layers being probed by the cooling right after the end of the outburst. Since our data are consistent with a range of values for the early-time power-law slope, we note that slopes in the range 0.01--0.08 give energies of $\simeq$4--32 keV per accreted nucleon. Heat deposits per nucleon in this range are consistent with the energies available from electron captures in the outer layers of the crust \citep{gupta2007,haensel2008}. We note that most of the {\it total} heat deposit per nucleon (which is of the order of an MeV; see below) is believed to be released in pycnonuclear fusion reactions much deeper in the crust, as mentioned in Section~\ref{sec:intro}.

The initial NS surface temperature for \sj\ in quiescence seems to be considerably higher than those of the other studied cooling sources. For \sj\ this temperature is $\simeq$165~eV, compared to $\simeq$110~eV for KS 1731--260, $\simeq$130~eV for MXB 1659--29, and $\simeq$120--130~eV for EXO 0748--676, as determined by extrapolating the best-fit exponential decay cooling curve back to the end of the outburst in each case. The high temperature for \sj\ possibly reflects the fact that it accreted at an extraordinarily high level during its outburst, much higher than any of the other sources, although it could also be related to a higher equilibrium temperature for \sj.
We note, however, that these values could be off by as much as $\simeq$10--20~eV due to uncertainties in the NS distance and radius, and exponential extrapolation. Moreover, it may not be appropriate to assume exponential decay at the start of quiescence. If the behavior was closer to a power law, the initial temperatures could be significantly higher for KS 1731--260, MXB 1659--29, and EXO 0748--676, but this is highly uncertain, due to the scarcity of early observations.

It is not clear whether the temperature of J1701 has reached its equilibrium value or thermal relaxation between the NS crust and core is still ongoing. The exponential decay fit to the cooling curve strongly indicates that the temperature has reached an approximately constant value. The situation is less clear when considering the broken power-law model (see Figure~\ref{fig:cooling_log}). The source could still be in a slow decay, or the power-law curve may already have flattened out as it is expected to eventually do. However, such flattening at perhaps $\sim$200 days post-outburst would have occurred much earlier than for KS 1731--260 and MXB 1659--29, where fits would indicate a value closer to $\sim$1000 days (BC09), and it is not even clear whether KS 1731--260 has stopped cooling. Further observations of \sj, 1000--2000 days post-outburst, are needed to confirm whether the source has reached equilibrium and, if not, to constrain the rest of the decay. If the crust has relaxed, then the equilibrium surface temperature of \sj\ is much higher than the values reported for KS 1731--260 and MXB 1659--29. The best-fit equilibrium temperature for \sj\ is $125.0\pm0.9$ eV, compared to $70.2\pm1.2$ eV for KS 1731--260 (or perhaps less, given the recent observation mentioned in Section~\ref{sec:intro}) and $54\pm2$ eV for MXB 1659--29 \citep{cackett2008}. Uncertainties in the NS distance and radius (as well as, in the case of \sj, uncertainties due to possible effects from the non-thermal component on the derived temperatures) make these values uncertain by $\simeq$5--15 eV. This implies a total temperature drop for \sj\ of $\simeq$40 eV, similar to that seen for KS 1731--260 (not taking into account the most recent observation), but considerably less than the $\simeq$75 eV drop for MXB 1659--29 (note, however, our caveat above about uncertainties in the initial temperatures). The equilibrium temperature is set by the temperature of the NS core, which in turn depends on the long-term time-averaged mass accretion rate of the NS and the extent to which the core is able to cool via neutrino emission. Unfortunately, all the information we have on the accretion history of \sj\ is the recent outburst and the fact that before the outburst the source had probably been in quiescence at least since the start of the \xte\ mission and its all-sky monitoring in 1996 January, although with the ASM we cannot rule out long-term activity at luminosities $\sim$$10^{34}$--$10^{36}\textrm{ erg s}^{-1}$ or short-term activity at higher luminosities. The recurrence time for outbursts in this system is therefore unknown, and we do not know whether the observed outburst is representative of typical behavior for the source. The possible equilibrium temperature of \sj\ corresponds to a bolometric thermal luminosity of $\simeq$$5.4\times10^{33}\textrm{ erg s}^{-1}$. This would be among the highest luminosities seen for a quiescent NS-LMXB, similar to those seen for Aql X-1 and 4U 1608--52 (see, e.g., \citealt{heinke2007,heinke2009}, and references therein).

To get some indication of how our results for \sj\ compare with theoretical predictions of the quiescent thermal luminosities of accreting NS transients, we use the results of \citet{yakovlev2004b}. They compute the quiescent bolometric thermal luminosity as a function of long-term time-averaged mass accretion rate for several models of accreting NSs warmed by deep crustal heating. They do this for several different equations of state, for NS masses between 1.1 $M_\sun$ and $\simeq$$2.0$ $M_\sun$, and for two models of a heavy-element accreted envelope: nuclear burning ashes composed of ${}^{56}$Fe with a total heat deposit of 1.45 MeV per accreted nucleon \citep{haensel1990}, and ${}^{106}$Pd ashes with a heat deposit of 1.12 MeV per nucleon \citep{haensel2003}. They also take into account a possible He layer on top of the heavy-element envelope. We note that \cite{haensel2008} have since revised the estimates of the heat deposits to $\simeq$1.9 MeV for ${}^{56}$Fe and $\simeq$1.5 MeV for ${}^{106}$Pd. We compare our results to the computed luminosity curves for a NS of mass 1.1 $M_\sun$ (the authors note that the curves are nearly identical for 1.3 $M_\sun$, and are insensitive to the assumed equation of state among those they consider), and for a heat deposit of 1.45 MeV per nucleon (see Figure~5 in \citealt{yakovlev2004b}). Given the quoted 15\% uncertainty in the best-estimate distance to \sj\ and various uncertainties arising from, e.g., the assumed values of the NS parameters, the choice and fitting of the spectral model, and the extrapolation of the thermal component outside the observed energy range (see Sections~\ref{sec:spectral_fitting} and \ref{sec:cooling_curves}), we consider a representative range of (3--$9)\times10^{33}\textrm{ erg s}^{-1}$ for our assumed equilibrium bolometric thermal luminosity. For the upper limit luminosity of $9\times10^{33}\textrm{ erg s}^{-1}$ and no He layer, we get an upper limit accretion rate of $\sim$$2\times10^{-9}\textrm{ }M_\sun\textrm{ yr}^{-1}$. For the lower limit luminosity of $3\times10^{33}\textrm{ erg s}^{-1}$ and the case of a thick He layer, we get a lower limit accretion rate of $\sim$$4\times10^{-11}\textrm{ }M_\sun\textrm{ yr}^{-1}$. This range is consistent with  constraints on the accretion rates of many NS transients (see, e.g., \citealt{heinke2007,heinke2009}). Assuming that the 2006--2007 outburst is typical for \sj, these long-term averaged accretion rates, in conjunction with our estimated average outburst rate (see discussion earlier in this section), give recurrence times for outbursts in the system ranging from $\sim$10 years to $\sim$700 years. Our data for \sj\ are therefore entirely consistent with standard cooling in a low-mass NS, and there is no need to assume a higher-mass NS with enhanced cooling. In fact, significantly enhanced cooling is somewhat unlikely, since in that case a very high long-term average accretion rate would be required to keep the NS as warm as it is observed to be; this suggests a rather low-mass NS in \sj, especially for NS models assuming nucleon or nucleon--hyperon matter in the core (see Figure~3 in \citealt{yakovlev2004b} and Figure~12 in \citealt{yakovlev2004}). We note that our estimate of the outburst accretion rate could well be off by a factor of $\sim$5--10 given uncertainties in the energy output of the outburst and the radiative efficiency of the accretion, and substantial uncertainties are of course associated with the theoretical calculations. Nevertheless, we conclude that if the crust of \sj\ has indeed already reached (or is close to reaching) thermal equilibrium with the core, then a rather low-mass NS with a near-standard cooling scenario is more likely than a high-mass star with significantly enhanced cooling.

\subsection{Behavior of the Non-thermal Component}\label{sec:non-thermal}

We speculate that the increase in temperature seen for the third \xmm\ observation (XMM-3) is due to an additional spurt of accretion. This is supported by the fact that XMM-3 has a large non-thermal flux, which is suggestive of ongoing accretion given the fact that prominent non-thermal components are commonly seen in spectra from accreting NS systems at low luminosity (see, e.g., \citealt{disalvo2002,barret2000}). During such an accretion spurt the surface of the NS may have been subjected to some shallow and temporary reheating. Furthermore, \citet{zampieri1995} show that low-level accretion onto a NS surface can produce radiation with a thermal (hardened blackbody-like) spectrum; additional thermal flux of this sort could also explain the increase in the inferred effective temperature. The subsequent \cxo\ observation (CXO-4) seems to have an anomalously high temperature as well, but has a much smaller non-thermal flux than XMM-3. As discussed in Section~\ref{sec:tests}, the high effective temperature derived for CXO-4 may be a result of the non-thermal component in the spectrum skewing our estimate of the thermal flux; a temperature in line with the overall observed decay is also consistent with the data. However, if the heightened thermal flux in CXO-4 is real and due to accretion, then the fact that this observation does not show a large thermal flux compared to the other observations (in contrast to XMM-3) is possibly indicative of some change in the properties of the accretion flow between XMM-3 and CXO-4. In any case, the substantial decrease in both the thermal and non-thermal flux compared to XMM-3 would point to a significant decrease in the rate of this possible accretion. We note that no activity was seen with the \xte\ ASM around the time of XMM-3 (or at any other time since the start of quiescence). This is hardly surprising, since an increase in luminosity of several orders of magnitude would be required for an ASM detection.

In general, a high degree of variability is seen in the non-thermal flux from \sj. Variability by a factor of $\simeq$30 ($\simeq$10 when excluding XMM-3) is observed between the individual observations obtained throughout the first $\simeq$800 days of the quiescent phase. Indications of possible variability on a timescale of $\sim$1--3 days is seen in the data from CXO-7 and CXO-10. As irregular variability on various timescales is a common characteristic of accretion, this points to accretion of some sort as a strong candidate for the cause of the non-thermal component. We searched for short-term variability (within observations) by constructing light curves for the \xmm\ observations, but did not see indications of variability. For each \cxo\ observation we performed a one-sided K--S test comparing photon arrival times to a uniform count rate model. This did not reveal any evidence for variability at a confidence above 90\%, with the exception of the first of the two exposures in CXO-10, which showed variability at a 97.7\% confidence level (discussed in Section~\ref{sec:tests}). However, due to the low count rates in our observations this does not set strong constraints on the possible existence of short-term variability.

There seems to be a significant difference between the value of the power-law index seen for XMM-3 ($\simeq$1.3--1.4) and those seen for the \sw\ and \xte\ observations during the transition to quiescence ($\simeq$1.7 and $\simeq$1.9 for the first and second \sw\ observations, and even higher for the third one, although highly uncertain due to very few counts; the \xte\ observations almost all had index values in the range $\simeq$1.8--2.5). Given that the outburst was still ongoing when the \sw\ and \xte\ observations were made, it is reasonable to assume that those spectra are dominated by flux due to accretion. This may indicate a different origin for the non-thermal component, or a difference in the accretion flow compared to the outburst, for XMM-3. We do note, however, that the luminosity in XMM-3 was lower, by a factor of 2 or more, than the luminosity in any of the \sw\ and \xte\ observations; the validity of a direct comparison is therefore questionable. The value of the combined power-law index for the other observations in quiescence is in line with those from the \sw\ and \xte\ observations, but this combined index does not give us information about the index values for individual observations. The fact that the combined index value seems to be different from the XMM-3 one may also point to some difference in nature between the non-thermal component in XMM-3 and that in the other quiescent observations.

A study by \citet{jonker2004} suggested that the fractional contribution of the non-thermal flux to the total unabsorbed 0.5--10 keV flux in non-pulsing NS-LMXB transients evolves as a function of total 0.5--10 keV luminosity. Based on data from several NS-LMXBs, \citet{jonker2004} suggest that the fraction decreases with decreasing luminosity down to a minimum of 10\%--20\% around $\sim$$2\times10^{33}\textrm{ erg s}^{-1}$, and increases again at lower luminosities to values of 60\%--70\% (see Figure~5 in their paper). In Figure~\ref{fig:pl_frac}, we show the fractional contribution of the power-law component to the total 0.5--10 keV flux as a function of the total luminosity in the same band. The power-law fraction varies rather randomly between $\simeq$5\% and $\simeq$50\%, and does not seem to exhibit behavior of the sort seen in the \citet{jonker2004} study.

\section{Summary}

We have presented \xte\ and \sw\ observations tracking the final 3 weeks of the 2006--2007 outburst of the super-Eddington NS transient XTE J1701--462, as well as \cxo\ and \xmm\ observations covering the first $\simeq$800 days of the subsequent quiescent phase. The source transitioned sharply from active accretion to quiescence, with the luminosity decreasing by a factor of $\sim$2000 in $\simeq$13 days. The end of the outburst is tightly constrained to a $\simeq$4 day window in 2007 August.

We fitted the \cxo\ and \xmm\ spectra with a two-component model consisting of a NS atmosphere model (thermal component, interpreted as being due to emission from the NS surface) and a power-law model (non-thermal component whose origin is uncertain). The effective surface temperature of the NS, inferred from the thermal component, was seen to decay rapidly in the first $\sim$200 days of quiescence. We interpret this as the NS crust cooling after having been heated and brought out of thermal equilibrium with the core during the outburst. The interpretation of the data is complicated by an increase in the derived temperature $\simeq$220 days into quiescence. The existence of the non-thermal component also adversely affects our ability to constrain the thermal component.

Fitting the derived temperatures with an exponential decay cooling curve plus a constant offset (excluding the two observations affected by the apparent temperature increase) we derive an $e$-folding time of $\simeq$$120^{+30}_{-20}$ days with a best-fit offset of $\simeq$125 eV. This baseline temperature is uncertain to $\simeq$15 eV due to uncertainties in the distance, radius, and mass of the NS, and in the modeling of the non-thermal component. The short decay timescale strongly indicates high thermal conductivity in the NS crust, and possibly suggests low-impurity material \citep{brown2009}.

The temperature data can also be well fitted with a more physically motivated broken power-law model \citep{brown2009}. The data show a strong indication for a break in the power law; the time of the break is $\simeq$20--150 days after the end of the outburst. This is considerably earlier than the break predicted by theory, which is mainly due to a change in heat capacity in the NS crust where the material transitions from a classical to a quantum crystal \citep{brown2009}. The observed break may therefore have a different origin; a possible alternative explanation is a strong nuclear heating source in the crust, e.g., the fusion of $^{24}$O \citep{horowitz2008}. The initial slope of the power-law cooling curve is a direct measure of the inward flux in the outer crust during outburst \citep{brown2009}. Our measured slope, in conjunction with an estimate for the accretion rate during the outburst, yields an estimate for the heat deposit per accreted nucleon in the outermost layers of the crust which is consistent with theoretical predictions for the energy available from electron captures \citep{gupta2007,haensel2008}.

Further observations are needed to determine whether the crust is still cooling slowly or has already reached thermal equilibrium with the core at a surface temperature of $\simeq$125 eV. The latter would imply an equilibrium bolometric thermal luminosity of $\simeq$$5\times10^{33}\textrm{ erg s}^{-1}$ for an assumed distance of 8.8 kpc. This would be among the highest quiescent thermal luminosities seen from a NS-LMXB, and may indicate a rather low-mass NS without significantly enhanced cooling.

The non-thermal component has varied irregularly throughout the quiescent phase by a factor of $\simeq$30; indications of possible variability have been seen on timescales as short as $\sim$1 day. The fractional contribution of this component to the total flux has been between $\simeq$$5\%$ and $\simeq$$50\%$. We speculate that the non-thermal component in XTE J1701--462 arises from residual accretion, and that the increase in derived temperature $\simeq$220 days into quiescence (which was accompanied by a large increase in non-thermal flux) was due to a spurt of increased accretion, possibly causing some shallow and temporary reheating of the NS surface and/or releasing radiation with a thermal spectrum \citep{zampieri1995}.

The observed behavior of XTE J1701--462 during the post-outburst quiescence seems to be quite different from that of other cooling NS transients observed after extended outbursts: the derived exponential decay timescale of the effective surface temperature is much shorter; both the initial and final surface temperatures are likely significantly higher; there is a strong indication for a break in the cooling curve when fitting the temperature data with a (broken) power law; a significant temporary increase in thermal flux was observed after more than 200 days of quiescence; finally, the source has exhibited a prominent non-thermal component in its spectrum throughout the quiescent phase.

\acknowledgements

This work was supported by \cxo\ Awards GO7-8049X and GO9-0057X. T.M.B.\ acknowledges support from ASI via contract I/088/06/0. We thank the referee for constructive comments which helped improve the paper. This research has made use of data obtained from the High Energy Astrophysics Science Archive Research Center (HEASARC), provided by NASA's Goddard Space Flight Center.

\bibliography{/Users/joelkf/Research/documents/papers/bibliography/references}

\end{document}